\documentclass[journal]{IEEEtran}
\IEEEoverridecommandlockouts
\usepackage{amsfonts}
\usepackage{amsmath}
\usepackage{amssymb}
\usepackage{urwchancal}
\usepackage{algorithm}
\usepackage{algorithmicx}
\usepackage{algpseudocode}
\usepackage{dsfont}
\usepackage{authblk}
\usepackage{bbm}
\usepackage{graphicx}
\usepackage{epstopdf}
\usepackage{subfigure}
\usepackage{stfloats}
\usepackage{eqnarray}
\usepackage{makecell}
\usepackage{multirow}
\usepackage{enumerate}
\usepackage{booktabs}
\usepackage{cite}
\usepackage{balance}
\usepackage{color}
\usepackage{bm}
\usepackage{mathtools}

\renewcommand{\IEEEQED}{\IEEEQEDopen}
\newcommand{\OO}[1]{{\cal O}\br{#1}}

\newcommand{\obj}{{\cal OBJ}}

\newcommand{\hav}{\mbox{\tiny HAV}}

\newcommand{\cir}{\mbox{\tiny CIR}}
\renewcommand{\sp}{\mbox{\tiny SP}}

\newcommand{\bs}{\mbox{\tiny BST}}
\newcommand{\sg}{\mbox{\tiny SG}}

\newcommand{\opt}{{\cal OPT}}
\newcommand{\inter}{\mbox{\tiny INTER}}
\newcommand{\intra}{\mbox{\tiny INTRA}}
\newcommand{\act}{\mbox{\tiny ACT}}
\newcommand{\sinr}{\mbox{SINR}}

\renewcommand{\vec}{\mbox{vec}}
\newcommand{\lpe}{\mbox{\tiny LPE}}

\renewcommand{\H}{\mbox{\tiny H}}
\renewcommand{\tt}{\mbox{\tiny T}}
\newcommand{\A}{\mbox{\tiny A}}
\newcommand{\U}{\mbox{\tiny U}}
\newcommand{\F}{\mbox{\tiny F}}
\newcommand{\E}{\mathds{E}}
\newcommand{\C}{\mathbb{C}}

\newcommand{\N}{{\cal N}}
\newcommand{\T}{{\cal T}}
\newcommand{\X}{{\cal X}}
\newcommand{\Y}{{\cal Y}}
\newcommand{\Z}{{\cal Z}}
\newcommand{\q}{\bm q}
\renewcommand{\r}{\bm r}
\newcommand{\s}{\bm s}
\renewcommand{\v}{\bm \nu}
\newcommand{\h}{\bm h}
\newcommand{\w}{\bm w}

\newcommand{\XI}{\bm \Xi}

\DeclarePairedDelimiter\autobracket{(}{)}
\DeclarePairedDelimiter\autosquare{[}{]}
\newcommand{\abs}[1]{\left|{#1}\right|}
\newcommand{\br}[1]{\autobracket*{#1}}
\newcommand{\sq}[1]{\autosquare*{#1}}
\newcommand{\cb}[1]{\left\{#1\right\}}
\newcommand{\norm}[1]{\left\|#1\right\|_{\F}}

\newcommand{\st}{\mbox{s.t.}}
\newcommand{\sumn}{\sum\limits_{n=1}^{N_m}}
\newcommand{\summ}{\sum\limits_{m=1}^M}
\newcommand{\suml}{\sum\limits_{l \in {\cal N}_m}}
\newcommand{\sumk}{\sum\limits_{k=0}^{K-1}}

\newcommand{\sumtk}{\sum\limits_{t_k \in \T_k}}

\newtheorem{remark}{\textbf{Remark}}
\newtheorem{proposition}{\textbf{Proposition}}

\begin{document}
\title{\Large Cross-Layer Scheduling and Beamforming in Smart-Grid Powered Cellular Networks With Heterogeneous Energy Coordination}

\author{
\mbox{\small Yanjie Dong,~\IEEEmembership{\small Student Member, IEEE},}
\mbox{\small Md. Jahangir Hossain,~\IEEEmembership{\small Senior Member, IEEE},}
\mbox{\small Julian Cheng,~\IEEEmembership{\small Senior Member, IEEE},}
\mbox{\small and Victor C. M. Leung,~\IEEEmembership{\small Fellow, IEEE}}
\thanks{
This work was supported in part by a UBC Four-Year Doctoral Fellowship, in part by the Natural Science and Engineering Research Council of Canada, and in part by the National Engineering Laboratory for Big Data System Computing Technology at Shenzhen University, China. This paper was presented in part at the IEEE ICC, Shanghai, P.R. China, May 20--24, 2019. 
(\emph{Corresponding author: Victor C. M. Leung.})}
\thanks{Y. Dong is with the Department of Electrical and Computer Engineering, The University of British Columbia, Vancouver, BC V6T 1Z4, Canada (email:ydong16@ece.ubc.ca).}
\thanks{M. J. Hossain and J. Cheng are with the School of Engineering, The University of British Columbia, Kelowna, BC V1V 1V7, Canada (email:\{jahangir.hossain, julian.cheng\}@ubc.ca).}
\thanks{V. C. M. Leung is with the College of Computer Science and Software Engineering, Shenzhen University, Shenzhen 518060, China, and the Department of Electrical and Computer Engineering, The University of British Columbia, Vancouver, BC \mbox{V6T 1Z4}, Canada (e-mail: vleung@ieee.org).}
}

\maketitle
\thispagestyle{empty} 
\pagestyle{empty}  
\begin{abstract}
User scheduling, beamforming and energy coordination are investigated in smart-grid powered cellular networks (SGPCNs), where the base stations are powered by a smart grid and natural renewable energy sources. Heterogeneous energy coordination is considered in SGPCNs, namely energy merchandizing with the smart grid and energy exchanging among the base stations.
A long-term grid-energy expenditure minimization problem with proportional-rate constraints is formulated for SGPCNs. Since user scheduling is coupled with the beamforming vectors, the formulated problem is challenging to handle via standard convex optimization methods. In practice, the beamforming vectors need to be updated over each slot according to the channel variations. User scheduling needs to be updated over several slots (frame) since the frequent scheduling of user equipment can cause reliability issues. Therefore, the Lyapunov optimization method is used to decouple the problem. A practical two-scale algorithm is proposed to schedule users at each frame, and obtain the beamforming vectors and amount of exchanged natural renewable energy at each slot. We prove that the proposed two-scale algorithm can asymptotically achieve the optimal solutions via tuning a control parameter. Numerical results verify the performance of the proposed two-scale algorithm.
\end{abstract}

\section*{Nomenclature}
\begin{tabbing}
  \textbf{Variables} \quad \= \textbf{Definitions} \\
  $M$ \> Number of BSTs \\
  $L$ \> Number of antennas at each BST \\
  $N_m$ \> Number of associated UEs in the $m$th BST \\
  $T$ \> Number of slots in each frame \\
  $\T_k$ \> Set of slots in the $k$th frame \\
  $\h_{m,n}\br{t_k}$ \> Channel-coefficient vector of  the $\br{m,n}$th \\\> access link \\
  $\omega_{m,n}$ \> Pathloss of the $\br{m,n}$th access link\\
  $a_{m,n}\sq{k}$ \> Scheduled UE indicator \\
  $y_{m,n}\br{t_k}$ \> Received signal of the $\br{m,n}$th UE \\
  $\w_{m,n}\br{t_k}$ \> Single-stream beamforming vector for  \\\>  the $\br{m,n}$th UE\\
  $z_{m,n}\br{t_k}$ \> AWGN at the $\br{m,n}$th UE \\
  $\sigma^2_{m,n}$ \> Power of the AWGN at the $\br{m,n}$th UE\\
  $\sinr_{m,n}$ \> Received SINR of the $\br{m,n}$th UE \\
  $I^{\intra}_{m,n}\br{t_k}$ \> Intra-cell interference received at \\\> the $\br{m,n}$th UE at the $t_k$th slot \\
  $I^{\inter}_{m,n}\br{t_k}$ \> Inter-cell interference received at \\\> the $\br{m,n}$th UE at the $t_k$th slot \\
  $r_{m,n}\br{t_k}$ \> Data rate of the $\br{m,n}$th UE at   the $t_k$th slot \\
  $P_m^{\bs}\br{t_k}$ \> Consumed power of the $m$th BST at   the $t_k$th slot \\
  $P_m^{\cir}$ \> Consumed circuit power of the $m$th BST \\
  $P_m^{\sp}$ \> Consumed power on baseband processing  \\\> of   the $m$th BST \\
  $\eta$ \> Power amplifier efficiency of the BSTs\\
  $\N_m^{\act}\sq{k}$ \> Set of scheduled UEs of the $m$th BST  \\\>  at the $k$th frame \\
  $\alpha_b$ and $\alpha_s$ \> Purchasing and selling prices of a unit power \\
  $\delta_m^l\br{t_k}$ \>  Power delivered from the $m$th BST to  \\\> the $l$th BST at the $t_k$th slot\\
  $\beta_m^l$ \>  Efficiency of local power line between  \\\> the $m$th BST and the $l$th BST\\
  $\N_m$ \> Neighbor BST of the $m$th BST \\
  $E_m^{\lpe}\br{t_k}$ \>  Net exchanged energy via the local power  lines \\\> of   the $m$th BST at the $t_k$th slot \\
  $G_m^{\sg}\br{t_k}$ \>  Grid-energy expenditure of the $m$th BST  \\\> at the $t_k$th slot\\
  $q_{m,n}^{\A}\br{t_k}$ \>  Backlog of the $\br{m,n}$th access queue  \\\> at the $t_k$th slot\\
  $r_{m,n}\br{t_k}$ \>  Data rate of the $\br{m,n}$th UE at the $t_k$th slot\\
  $\nu_{m,n}\br{t_k}$ \>  Arrival rate of the $\br{m,n}$th UE at the $t_k$th slot\\
  $q^{\U}_{m,n}\br{t_k}$ \>  Backlog of the $\br{m,n}$th processing queue  \\\> at the $t_k$th slot \\
  $s_{m,n}\br{t_k}$ \>  Processing rate of the $\br{m,n}$th processing queue  \\\> at the $t_k$th slot \\
  $\tilde s_{m,n}$ \> Constant processing rate of the $\br{m,n}$th  \\\> processing queue \\
  $P_m^{\max}$ \> Maximum transmit power of the $m$th BST \\
  $\nu^{\max}$ \> Maximum arrival rate \\
  $r^{\max}$ \> Maximum data rate \\
  $s^{\max}$ \> Maximum service rate \\
  $\bar\nu_{m,n}$ \> Average arrival rate of the $\br{m,n}$th access queue \\
  $\bar s_{m,n}$ \> Average service rate of the $\br{m,n}$th  \\\> processing queue \\
  $\bar G$ and $G^*$ \>  Bound of grid-energy expenditure and  \\\> optimal grid-energy expenditure\\
  $d_{m,n}$ \> Link distance of the $\br{m,n}$th access link \\
  $f_c$ \> Carrier frequency
\end{tabbing}

\section{Introduction}
Wireless data is estimated to exceed $131$ exabytes per month by 2024 \cite{ericsson_outlook}.
Deploying ultra-dense base stations (BSTs) is a promising solution to cope with the ever-increasing volume of wireless data \cite{LiuSept.2018}.
One study estimates that radio access links will consume around 29\% of the energy consumed by wireless communications \cite{FehskeAug.2011, WuAug.2017}.
Hence, the operation of a large number of BSTs has led to surging energy bills and carbon footprint in the information and communication technology sector \cite{FehskeAug.2011}.
To reduce the energy bills of BSTs, research on green communications has attracted much attention from both industry and academia \cite{FehskeAug.2011, WuAug.2017, Suntobepublished}.

\subsection{Related Works and Motivations}
The current research on green communications can be classified into two categories.
The first category focuses on reducing energy consumption \cite{DahroujMay2010, LakshminarayanaOct.2015} or increasing energy efficiency \cite{HeJune2014, LiNov.2014}.
For example, the energy consumption minimization problems with short-term and long-term constraints on communication quality of service (QoS) were respectively studied in \cite{DahroujMay2010} and \cite{LakshminarayanaOct.2015} for downlink beamforming of multicell networks.
Downlink beamforming and downlink power control were, respectively, investigated to maximize the short-term energy efficiency \cite{HeJune2014} and long-term energy efficiency \cite{LiNov.2014} of multicell networks.

Another research direction leverages the integration of natural renewable energy (NRE) into the cellular networks as energy-harvesting equipment (e.g., residential-level solar photovoltaic panels and miniature wind turbines) becomes widely available.
For example, Huawei and Telefonica have installed solar-powered BSTs in central Chile \cite{huawei_green}.
Several works investigated resource-allocation algorithms for communication systems powered by NRE and traditional grids (i.e., hybrid-powered communication systems), such as
point-to-point systems \cite{KangAug.2014, HuAug.2015, CuiAug.2015},
multiuser systems \cite{NgJul.2013, ZhaiNov.2015} and
multicell networks \cite{MaoDec.2015, Wei2018, GuoJuly2018}.
In the point-to-point systems, the joint allocation of harvested NRE and grid energy under the constraints of communication QoS was investigated to minimize the system cost and grid energy expenditure over finite-time horizon \cite{KangAug.2014, HuAug.2015} and infinite-time horizon \cite{CuiAug.2015}.
Moreover, the proposed algorithm in \cite{CuiAug.2015} achieves a tradeoff between the grid-energy expenditure and delay for the point-to-point systems.
Considering imperfect NRE storage media, an energy-delay tradeoff was also revealed for the hybrid-powered multiuser systems \cite{ZhaiNov.2015}.
However, the aforementioned works \cite{KangAug.2014, HuAug.2015, CuiAug.2015, NgJul.2013, ZhaiNov.2015, MaoDec.2015, Wei2018, GuoJuly2018} focused on the traditional grids and ignore a key feature of smart grids: two-way energy-trading capability \cite{LeeJuly2014}.
Two-way energy-trading provides another dimension to reduce the energy bills of hybrid-powered BSTs.
Hence, incorporating NRE into the smart-grid powered cellular networks (SGPCNs) becomes an ecologically and economically friendly solution to reduce the energy bills.

Several works have investigated the framework of SGPCNs \cite{WangNov.2016, BuAug.2012, XuJune2015,  FarooqFeb.2017} and the applicable resource allocation algorithms \cite{Xu2015, HuangJune2017, ShengFeb.2017, HanAug.2013, HuAug.2016, DongDec.2017, WangMay2016}.
More specifically, resource allocation algorithms in SGPCNs can be classified into three categories: one-shot algorithms, offline algorithms and online algorithms. 
The one-shot algorithms proposed in \cite{Xu2015, HuangJune2017, ShengFeb.2017} are applicable to scenarios where the resources are independently allocated in each slot of an SGPCN.
The offline algorithms were proposed to allocate jointly the SGPCN resources over a finite number of slots \cite{HanAug.2013, HuAug.2016}, whereas the online algorithms were tailored to handle the volatility of NRE arrivals in an SGPCN over an infinite number of slots \cite{DongDec.2017, WangMay2016, 7887725}.
Since practical SGPCNs operate in infinite time horizons, the online resource allocation algorithms are preferred.
For example, Dong \emph{et al.} \cite{DongDec.2017} investigated the impact of volatility of NRE arrivals on the packet rates.
Studying a long-term grid-energy-expenditure (LTGEE) minimization problem, they revealed that the grid-energy expenditure can be reduced by sacrificing the system packet rate.
Wang \emph{et al.} \cite{WangMay2016} studied the long-term data-rate maximization problem with a constraint on the long-term grid-energy expenditure in smart-grid powered communications.
Using the dirty paper coding scheme at the multiple-input-multiple-output BST, their proposed online beamforming algorithm \cite{WangMay2016} can achieve the provable asymptotically optimal data rate.
The authors in \cite{7887725} investigated the joint beamforming and grid-energy merchandizing problem to minimize the long-term grid-energy expenditure  in a coordinated multi-point system based on the group-sparse optimization method and the multi-armed bandit method.
Since the formulated long-term grid-energy expenditure is a function of ahead-of-time energy-trading amount, the relation between the long-term grid-energy expenditure and the beamforming vectors is unknown in \cite{7887725}.

The  online resource allocation algorithms \cite{WangMay2016, DongDec.2017} reviewed above allocate resources over one time scale.
The scheduled user equipment (UE) indicators need to be reallocated over several slots in practical systems since the frequent scheduling of the UEs can cause reliability issues.
Moreover, when the UEs can tolerate the delay, the frame-scale user scheduling induces a more accurate characterization of end-to-end delay.
Few studies have investigated two-scale resource allocation schemes.
Wang \emph{et al.}  proposed in \cite{WangMay2018} the dynamic beamforming and grid-energy merchandizing algorithm to minimize the long-term grid-energy expenditure for the multiuser SGPCNs.
Their proposed two-scale algorithm \cite{WangMay2018} allocates: 1) the ahead-of-time energy-trading amount at frame scale; and 2) the real-time energy-trading amount and beamforming at the slot scale. 
Since the ahead-of-time energy-trading amount is a continuous variable, it can be obtained by the subgradient method.
Since the scheduled UE indictors are binary variables, the proposed schemes in \cite{WangMay2016, DongDec.2017, WangMay2018, 7887725} cannot obtain the optimal scheduled UE indicators.  ~Yu \emph{et al.} \cite{YuApr.2016}  investigated  the joint network selection, subchannel and power allocation problem in integrated cellular and \mbox{Wi-Fi} networks.
Exhaustive search is used in \cite{YuApr.2016}  to solve the network selection subproblem, and greedy selection is used to solve the subchannel allocation subproblem.
However, exhaustive search is computational expensive, and greedy selection can lead to suboptimal solutions for the scheduled UE indicators when the number of UEs is large.
Besides, the proposed algorithms in \cite{WangMay2018, YuApr.2016} are not applicable when  heterogeneous energy coordination and proportional-rate constraints are considered in  multicell SGPCNs.
In our previous work \cite{Dongtobe}, we proposed a joint scheduling and beamforming algorithm to minimize the long-term grid-energy expenditure.
A tradeoff  between the grid-energy expenditure and the end-to-end delay of UEs was unveiled.
However, the tradeoff is yet to be found for SGPCNs having heterogeneous energy coordination.
Besides, a theoretical analysis on the tradeoff was not performed in \cite{Dongtobe}.
Compared with the Lyapunov optimization methods used in \cite{YuApr.2016, WangMay2016, DongDec.2017, WangMay2018, LiMar.2019},  reinforcement learning methods as applied in \cite{Wei2018, SadeghiAug.2019} can also be used to develop online optimization algorithms.
However,  reinforcement learning  requires the proper development of a function estimator to deal with the continuous states and continuous actions.
Besides, it is more challenging to handle the constrained optimization problems via the reinforcement learning approach.
Therefore, we are motivated to adopt the Lyapunov optimization method to solve the two-scale resource allocation problem.

\subsection{Contributions}
Different from \cite{WangMay2016, DongDec.2017, WangMay2018}, we consider the heterogeneous energy coordination in  SGPCNs such that each BST has two options to handle the harvested NRE: 1) energy merchandizing with the smart grid; and 2) energy exchanging with the other BSTs.
We investigate the LTGEE minimization problem in a SGPCN via the joint allocation of scheduled UE indicators, beamforming and exchanged NRE variables.
By tuning a control parameter of the Lyapunov optimization method, we can asymptotically obtain the optimal grid-energy expenditure.
The contributions of this work are summarized as follows.
\begin{itemize}
  \item We investigate the LTGEE minimization problem in  SGPCNs via the joint design of scheduled UE indicators, beamforming vectors, and exchanged NRE variables.
      The design of beamforming vectors and exchanged NRE variables belongs to the physical layer, and the design of scheduled UE indicators belongs to the link layer.
      Hence, the investigated LTGEE minimization problem is a cross-layer problem.
  \item After transforming the LTGEE minimization problem into minimizing the upper bound of drift-plus-penalty function, we decouple the beamforming vectors and exchanged NRE variables from the scheduled UE indicators.
      Hence, a two-scale UE scheduling, beamforming and energy trading (TSUBE) algorithm is proposed to update the beamforming vectors and exchanged NRE variables every slot and update the scheduled UE indicator every frame.
  \item We theoretically prove that the minimizer to the upper bound of drift-plus-penalty function can be obtained via the proposed TSUBE algorithm.
      Based on the Lyapunov optimization method, we reveal that the proposed TSUBE algorithm can approach the optimal grid-energy expenditure via tuning a control parameter.
      Compared with the state-of-the-art analysis methods \cite{Neelybook, YuApr.2016, DongDec.2017, WuNov.2018, ZhaiJune2018}, the proposed method explicitly specifies the stable region of the SGPCN.
\end{itemize}


\subsection{Organization and Notations}
The remainder of this paper is organized as follows.
The system model and the LTGEE minimization problem are presented in Section II.
The tradeoff between the long-term \mbox{grid-energy} expenditure and the \mbox{end-to-end} delay of UEs is theoretically established in Section III.
Three dimensional resources (scheduled UE indicators, beamforming vectors and exchanged NRE variables) are optimally allocated to minimize the long-term grid-energy expenditure in Section IV.
Numerical results are presented to verify the effectiveness of the proposed TSUBE algorithm and insights from the results are discussed in Section V.
Finally, Section VI concludes the work.

\emph{Notations:}
Vectors and matrices are shown in bold lowercase letters and bold uppercase letters, respectively.
$\mathbb{C}$ denotes the domain of complex values.
${\cal CN}\br{\bm\mu, \bm\Sigma}$ denotes a circularly symmetric complex,
Gaussian (CSCG) random vector with mean vector $\bm\mu$ and covariance matrix $\bm\Sigma$.
The operator $\vec\br{\cdot}$ converts an $M\times N$ matrix into a column vector of size $MN\times 1$.
$\norm{\cdot}$ denotes the Frobenius norm.
The expectation of a random variable is denoted by $\E\cb{\cdot}$, and the imaginary part of a complex value is denoted by $\Im\br{\cdot}$.
The operators $\br{\cdot}^{\tt}$ and $\br{\cdot}^{\H}$ denote the transpose  and conjugate-transpose operations, respectively.
The vectors $\bm 1$ and $\bm 0$ denote the all-one and all-zero vectors, respectively.

\section{System Model and Problem Formulation}
\begin{figure}[htb]
\centering
  \includegraphics[width=3.3 in]{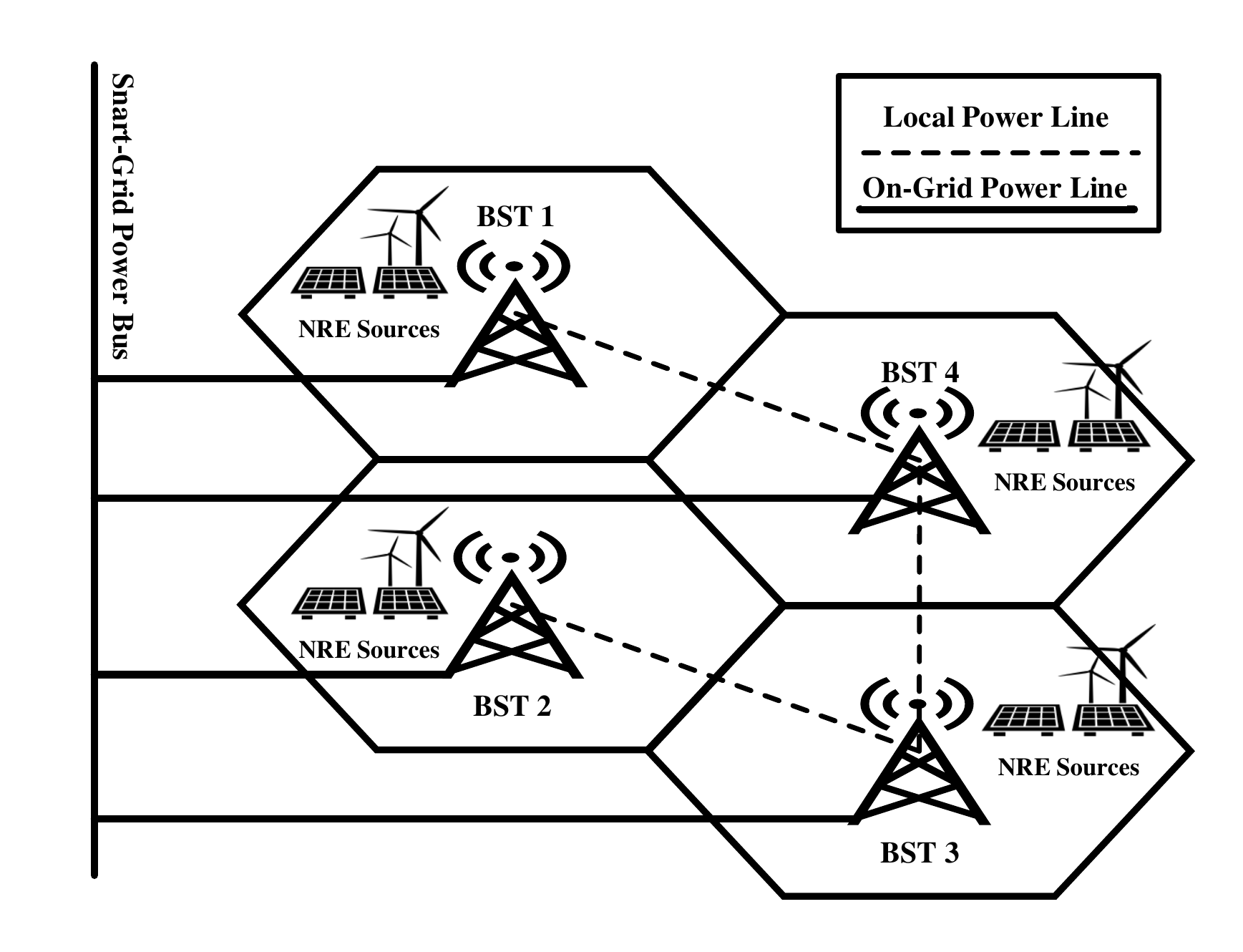}\\
  \caption{An illustration of the green cellular network with heterogeneous energy coordination.
  Each BST is powered by a smart grid and NRE sources.
  Therefore, the BSTs exchange NRE via local physical power lines or merchandize NRE via on-grid power lines.}\label{fg:001}
\end{figure}

As shown in Fig. \ref{fg:001}, we consider a SGPCN having $M$ BSTs.
Each BST is equipped with $L$ antennas.
The \mbox{$m$th} BST is associated with $N_m$ UEs, each equipped with a single antenna.
Each BST connects to the core network (CN) via optical-fiber links, and connects to UEs via the wireless links.
Moreover, each BST is powered by the smart grid and NRE sources (e.g., solar and/or wind).
A two-scale framework is considered to schedule UEs, design beamforming vectors and exchange NRE.
Since the NRE arrival rates and channel-coefficient vectors vary at different time scales in practice \cite{LiApr.2015}, we assume that the average NRE arrival rate is updated per frame, and the channel-coefficient vector is updated per slot.
Here, each frame consists of $T$ consecutive slots.
We denote the range of the frames by $k = 0, 1, \ldots, \infty$, and denote the range of slots in the $k$th frame by $t_k \in \T_k \triangleq \cb{t_k \left|t_k = kT, kT+1, \ldots, \br{k+1}T - 1\right.}$.
Moreover, we assume that each slot has unit duration; therefore, we use the terms ``energy'' and ``power'' interchangeably at the time scale of a slot.

\subsection{Signal Model}
Let $\bm h_{m,n}\br{t_k} \in \C^{L\times 1}$ denote the channel-coefficient vector of the link between the $n$th UE and the \mbox{$m$th} BST (or the $\br{m,n}$th access link) at the $t_k$th slot.
Here, $\bm h_{m,n}\br{t_k} \sim {\cal CN}\br{\bm 0, \omega_{m,n}^{-1}\bm I_{L}}$ follows CSCG where $\omega_{m,n}$ is the pathloss of the $\br{m,n}$th access link.
We define the scheduled UE indicator as $a_{m,n}\sq{k}$ which equals one when the $n$th UE of the $m$th BST (or the $\br{m,n}$th UE) is scheduled at the \mbox{$k$th} frame; otherwise, it equals zero. Hence, the received signal of the $\br{m,n}$th UE at the $t_k$th slot is given by\footnote{Here, the received signal $y_{m,n}\br{t_k}$ is a discrete-time signal since $t_k \in \T_k \triangleq \cb{t_k \left|t_k = kT, kT+1, \ldots, \br{k+1}T - 1\right.}$. However, we have slightly abused the notation to denote a signal at the slot scale.}
\begin{align}
y_{m, n}\br{t_k} =&  \sqrt{a_{m,n}\sq{k}} \bm h_{m,n}^{\H}\br{t_k}\bm w_{m,n}\br{t_k} \label{eqa:01}\\
& + \sum\limits_{i \neq n} \sqrt{a_{m,i}\sq{k}} \bm h_{m,n}^{\H}\br{t_k}\bm w_{m,i}\br{t_k} \nonumber\\
& + \sum\limits_{j \neq m}\sum\limits_{i = 1}^{N_j} \sqrt{ a_{j,n}\sq{k} } \bm h_{j,n}^{\H}\br{t_k}\bm w_{j,i}\br{t_k}  + z_{m,n}\br{t_k}. \nonumber
\end{align}
where the term $z_{m,n}\br{t_k} \sim {\cal CN}\br{0, \sigma_{m,n}^2}$ is the additive white Gaussian noise (AWGN) of the $\br{m, n}$th UE at the $t_k$th slot;
$\bm w_{m,n}\br{t_k}$ denotes the single-stream beamforming vector for the $\br{m,n}$th UE at the $t_k$th slot.

Based on \eqref{eqa:01}, the received signal-to-interference-plus-noise ratio (SINR)
of the $\br{m,n}$th UE at the $t_k$th slot is obtained  as
\begin{equation}\label{eqa:02}
\sinr_{m,n}\br{t_k} = \frac{ a_{m,n}\sq{k} \left|\bm h_{m,n}^{\H}\br{t_k}\bm w_{m,n}\br{t_k}\right|^2}{I_{m,n}^{\intra}\br{t_k} + I_{m,n}^{\inter}\br{t_k} + \sigma_{m,n}^2}
\end{equation}
where the intra-cell interference and inter-cell interference terms are, respectively, given as
\begin{equation}\label{eqa:03}
I_{m,n}^{\intra}\br{t_k} = \sum\limits_{i\neq n}  {a_{m,i}\sq{k}} \left|\bm h_{m,n}^{\H}\br{t_k}\bm w_{m,i}\br{t_k}\right|^2
\end{equation}
and
\begin{equation}\label{eqa:04}
I_{m,n}^{\inter}\br{t_k} = \sum\limits_{j \neq m}\sum\limits_{i = 1}^{N_j} {a_{j,i}\sq{k}} \left|\bm h_{j,n}^{\H}\br{t_k}\bm w_{j,i}\br{t_k}\right|^2.
\end{equation}
Hence, the data rate of the $\br{m,n}$th access link at the $t_k$th slot is given as

\noindent  $r_{m,n}\br{t_k} = \log\br{1 + \sinr_{m,n}\br{t_k}}$.

Based on \eqref{eqa:01}, the consumed power of the $m$th BST at the $t_k$th slot is written as
\begin{equation}\label{eqa:05}
P^{\bs}_m\br{t_k}
= \frac{1}{\eta} \sum\limits_{n \in \N_m^{\act}\sq{k}} \left\|\bm w_{m,n}\br{t_k}\right\|_{\F}^2 +  P_m^{\cir}
\end{equation}
where $\eta$ is the power amplifier efficiency of the BST, and the circuit power consumption is defined as $P_m^{\cir} \triangleq P_m^{\sp}\br{0.87 + 0.1 L + 0.03 L^2}$, in which $P_m^{\sp}$ is the power consumed by baseband processing in the \mbox{$m$th} BST \cite{DongDec.2017}.
Here, ${\cal N}_m^{\act}\sq{k}$ denotes the set of scheduled UEs of the $m$th BST at the $k$th frame.

\begin{remark}
Different from the conference version \cite{Dongtobe}, in this paper we do not consider on-off switching of the BSTs.
This is because in our numerical experiments, we observed that on-off switching of a BST rarely occurs when the BST has three or more UEs attached.
\end{remark}


\subsection{Energy-Coordination Model}
As shown in Fig. \ref{fg:001}, the BSTs have two options to perform heterogeneous energy coordination: 1) energy merchandizing via the on-grid power lines; and 2) energy exchanging via the local power lines.

\subsubsection{Energy Merchandizing}
Using a smart meter, a BST can  trade energy bidirectionally with the smart grid by purchasing (selling) energy when the NRE of the BST has a deficit (surplus).
Denote the unit energy purchasing price of the BSTs by $\alpha_b$, and the unit energy selling price of the BSTs by $\alpha_s$.
To avoid the redundant energy merchandizing, we set $\alpha_b > \alpha_s \ge 0$.

\subsubsection{Energy Exchange}
BSTs that are connected by local power lines can exchange energy exchange among themselves.
Due to the issues of regulation and resistive loss, the BSTs are partially connected\footnote{The BSTs are called partially connected when there exists at least one pair of BSTs that are not connected by a local power line in one hop.} as shown in Fig. \ref{fg:001}.
Let $\delta_{m}^{l}\br{t_k}$ and $\delta_{l}^{m}\br{t_k}$ respectively denote the energy delivered from the \mbox{$m$th} BST to the \mbox{$l$th} BST and the reversed direction at the $t_k$th slot.
Two-way energy flow needs to be avoided at a specific slot.
Hence, we include the energy-flow constraints as
\begin{equation}\label{eqa:06}
\delta_{m}^{l}\br{t_k} + \delta_{l}^{m}\br{t_k} = 0, \forall m, l.
\end{equation}

The case $\delta_{m}^{l}\br{t_k} \ge 0$ indicates that NRE is delivered from the \mbox{$m$th} BST to the \mbox{$l$th} BST, and vice versa.
Moreover, $\delta_m^m\br{t_k} = 0$, $m = 1, 2, \ldots, M$.

We consider the loss over the local power lines by defining the efficiency of power delivery over the local power line between the $m$th BST and the $l$th BST as $\beta_m^l \in \br{0, 1}$ with $\beta_m^l = \beta_l^m$.
Let the set $\N_m$ be the neighbor BSTs that are connected by one-hop local power lines to the $m$th BST.
The amount of net exchanged energy via the local power lines of the $m$th BST at the $t_k$th slot is written as
\begin{equation}\label{eqa:07}
E_m^{\lpe}\br{t_k} = \suml\max\cb{\delta_m^l\br{t_k}, \beta_m^l\delta_m^l\br{t_k}}.
\end{equation}

\begin{remark}
The amount of net exchanged energy in \eqref{eqa:07} is calculated as follows.
\begin{itemize}
  \item When $\delta_{m}^{l}\br{t_k} > 0$, energy is flowing from the $m$th BST to the $l$th BST in the $t_k$th slot, with 
      net output energy of the $m$th BST and net input energy of the $l$th BST being, respectively, $\delta_m^l\br{t_k}$ and $\beta_m^l\delta_m^l\br{t_k}$.
      Therefore, the net output energy of the $l$th BST is $-\beta_m^l\delta_m^l\br{t_k}$.
      Based on the energy-flow constraints in \eqref{eqa:06} and the fact $\beta_m^l = \beta_l^m$, the net output energy of the $l$th BST is $\beta_l^m\delta_l^m\br{t_k}$.
  \item When $\delta_{m}^{l}\br{t_k} < 0$, energy is flowing from the $l$th BST to the $m$th BST in the $t_k$th slot.
      Following similar arguments, we obtain the net energy output at the $m$th BST as $\beta_m^l\delta_m^l\br{t_k}$, and obtain the net energy output at the $l$th BST as $\delta_l^m\br{t_k}$.
\end{itemize}
\end{remark}

When energy merchandizing and energy exchanging are deployed, the grid-energy expenditure of the $m$th BST at the $t_k$th slot is calculated as \cite{DongDec.2017}
\begin{equation}\label{eqa:08}
\begin{split}
 G_m^{\sg}\br{t_k}
=& \br{\alpha_b - \alpha_s}\br{ P_m^{\bs}\br{t_k} + E^{\lpe}_m\br{t_k} - \frac{1}{T}E^{\hav}_m\sq{k} }^+ \\
& + \alpha_s\br{ P_m^{\bs}\br{t_k} + E^{\lpe}_m\br{t_k} - \frac{1}{T}E^{\hav}_m\sq{k} }
\end{split}
\end{equation}
where $E_m^{\hav}\sq{k}$ denotes the amount of  NRE harvested by the $m$th BST in the $k$th frame.
Since the arrival of NRE remains stable in a frame, the amount of  NRE harvested by the $m$th BST is $\frac{1}{T}E^{\hav}_m\sq{k}$.

\subsection{Traffic Model}
\subsubsection{Access Queue}
We consider that the $m$th BST maintains $N_m$ access queues for the associated UEs, and the dynamic equation for the $n$th access queue of the $m$th BST (or the $\br{m,n}$th access queue) is given as
\begin{equation}\label{eqa:09}
q_{m,n}^{\A}\br{t_k+1} = q_{m,n}^{\A}\br{t_k} - r_{m, n}\br{t_k} + \nu_{m, n}\br{t_k}
\end{equation}
where $q_{m,n}^{\A}\br{t_k+1}$ and $q_{m,n}^{\A}\br{t_k}$ are the backlogs of the $\br{m,n}$th access queue at the \mbox{$\br{t_k +1}$th} and \mbox{$t_k$th} slots, respectively;
$\nu_{m,n}\br{t_k}$ and $r_{m,n}\br{t_k}$ are, respectively, the arrival rate and data rate of the \mbox{$\br{m,n}$th} access queue at the $t_k$th slot.

\subsubsection{Processing Queue}
Corresponding to the $\br{m,n}$th access queue, we consider that the $\br{m,n}$th UE maintains a processing queue (or the \mbox{$\br{m,n}$th} processing queue) for upper layer processing.
The dynamic equation for the $\br{m,n}$th processing queue is given as
\begin{equation}\label{eqa:10}
q_{m,n}^{\U}\br{t_k + 1} = q_{m,n}^{\U}\br{t_k} - s_{m,n}\br{t_k} + r_{m,n}\br{t_k}
\end{equation}
where $q_{m,n}^{\U}\br{t_k + 1}$ and $q_{m,n}^{\U}\br{t_k}$ are the backlogs at  the $\br{t_k+1}$th and the $t_k$th slots.
We consider that the processing rate of the $\br{m,n}$th processing queue is constant.
Therefore, the processing rate $s_{m,n}\br{t_k} \triangleq \min\cb{\tilde s_{m,n}, q_{m,n}^{\U}\br{t_k}}$ where $\tilde s_{m,n}$ denotes the constant processing rate of the $\br{m,n}$th processing queue.

In practical systems, the values of arrival rate $\nu_{m,n}\br{t_k}$, data rate $r_{m,n}\br{t_k}$, and processing rate $s_{m,n}\br{t_k}$ are bounded as
\begin{equation}\label{eqa:11}
\begin{split}
\nu_{m,n}\br{t_k} &\in \sq{0, \nu^{\max}} \\
r_{m,n}\br{t_k} &\in \sq{0, r^{\max}} \\
s_{m,n}\br{t_k} &\in \sq{0, s^{\max}}
\end{split}
\end{equation}
where $\nu^{\max}$, $r^{\max}$ and $s^{\max}$ are, respectively, the maximum arrival rate, maximum data rate and maximum processing rate.

\begin{remark}
The arrival rate vector, data rate vector and processing rate vector are, respectively, denoted by $\v\br{t_k} \triangleq \vec( [\nu_{m,n}\br{t_k}]_{\forall m,n} )$, $\r\br{t_k} \triangleq \vec( [r_{m,n}\br{t_k}]_{\forall m,n} )$ and  $\s\br{t_k} \triangleq \vec( [s_{m,n}\br{t_k}]_{\forall m,n} )$.
The average arrival rate of the $\br{m,n}$th access queue and the average processing rate of the $\br{m,n}$th processing queue are, respectively, denoted by $\bar
\nu_{m,n} \triangleq \E_{\X}\cb{\nu_{m,n}\br{t_k}}$ and $\bar s_{m,n} \triangleq \E_{\X}\cb{s_{m,n}\br{t_k}}$.
Here, the operator $\E_{\X}\cb{\cdot}$ is the expectation over the random sources ${\X} \triangleq \cb{\X\br{t_k}}_{\forall t_k, k}$ where $\X\br{t_k} = \cb{\bm h_{m,n}\br{t_k}, E_m^{\hav}\sq{k}, \nu_{m,n}\br{t_k}}_{\forall m, n}$.
Moreover, the average arrival rate vector and average processing rate vector are denoted by $\bar\v \triangleq \vec([\bar\nu_{m,n}]_{\forall m,n})$ and $\bar\s \triangleq  \vec([\bar s_{m,n}]_{\forall m,n})$, respectively.
\end{remark}

\subsection{Problem Formulation}
Our objective is to minimize the long-term grid-energy expenditure via designing the scheduled UE indicators $\{a_{m,n}\sq{k}\}_{\forall m, n, k}$, beamforming vectors $\{\bm w_{m,n}\br{t_k}\}_{\forall l, m, n, t_k, k}$ and exchanged NRE variables $\{\delta_m^l\br{t_k}\}_{\forall l, m, t_k, k}$ over two time scales.
Due to the lack of knowledge on stochastic arrival of NRE and the variations of channel states in the future slots, we consider the following constraints in the LTGEE minimization problem.
\begin{itemize}
  \item Rate-limit constraints:
  \begin{equation}\label{eqa:12}
  r_{m, n}\br{t_k} \le q_{m,n}^{\A}\br{t_k}, \forall m,n
  \end{equation}
  which guarantee that each BST does not transmit blank information at the $t_k$th slot.
  \item Dynamic proportional-rate constraints:
  \begin{equation}\label{eqa:13}
  \frac{r_{m, n}\br{t_k}}{ r_{j, i}\br{t_k} } = \frac{q^{\A}_{m,n}\br{t_k}}{q^{\A}_{j,i}\br{t_k}}, n \in {\cal N}_m^{\act}\sq{k}, i \in {\cal N}_j^{\act}\sq{k}, \forall m, j
  \end{equation}
  which  guarantee that the UE with larger backlog obtains a better service rate at each slot.
  \item Slot-level power constraints:
  \begin{equation}\label{eqa:14}
  \sum\limits_{n \in {\cal N}_{m}^{\act}\sq{k}}  \left\|\bm w_{m,n}\br{t_k}\right\|_{\F}^2 \le P_m^{\max}, \forall m
  \end{equation}
  where $P^{\max}_m$ is the maximum transmit power of the $m$th BST.
  \item Queue-stable constraints:
  \begin{equation}\label{eqa:15}
  \limsup\limits_{K\rightarrow \infty}\frac{1}{K} \sumk \mathds{E}_{\cal X}\cb{ q_{m,n}^{\A}\sq{k} +  q_{m,n}^{\U}\sq{k}} < \infty, \forall m, n
  \end{equation}
  which guarantee that the data of UEs will be served in finite time.
\end{itemize}

As a result, the LTGEE minimization problem is formulated as
\begin{subequations}\label{eqa:16}
\begin{align}
\min\limits_{\Y} & \lim\limits_{K \rightarrow \infty} \frac{1}{KT}\sumk \sumtk \summ \mathds{E}_{\X}\cb{ G_m^{\sg}\br{t_k}} \label{eqa:16a}\\
\st\; &  \eqref{eqa:06}  \mbox{ and }  \eqref{eqa:12}-\eqref{eqa:15}
\end{align}
\end{subequations}
where $\Y \triangleq \cb{\Y\br{t_k}}_{\forall t_k, k}$ is the set of resource allocation variables, and the set $\Y\br{t_k}$ is defined as  $\Y\br{t_k} = \{\w_{m,n}\br{t_k}, \delta_m^l\br{t_k},  a_{m,n}\sq{k}\}_{\forall l, m, n}$.

Note that the LTGEE minimization problem \eqref{eqa:16} is challenging to handle via classical convex optimization methods.
Since the scheduled UE indicators are coupled with the beamforming vectors, we are motivated to use the Lyapunov optimization method to obtain a feasible solution to the LTGEE minimization problem \eqref{eqa:16}, analyze the optimality of the feasible solution. 
Moreover, we also investigate the relation between the long-term grid-energy expenditure and the end-to-end delay of UEs.

\section{Tradeoff Between Grid-Energy Expenditure and End-to-End Delay}
We define the Lyapunov function of LTGEE  minimization problem \eqref{eqa:16} at the $k$th frame as
\begin{equation}\label{eqa:17}
L\sq{k} =
\frac{1}{2} \norm{\q^{\A}\sq{k}}^2
+ \frac{1}{2}\norm{\q^{\U}\sq{k}}^2
\end{equation}
where $\q^{\A}\sq{k}$ is obtained by stacking the backlogs of the access queues, and $\q^{\U}\sq{k}$ is obtained by the backlogs of the processing queues at the $k$th frame as
$\q^{\A}\sq{k} \triangleq \vec({ [{q_{m,n}^{\A}\sq{k}}]_{\forall m, n} })$ and $\q^{\U}\sq{k} \triangleq \vec({ [{q_{m,n}^{\U}\sq{k}}]_{\forall m, n} })$, where
$q^{\A}_{m,n}\sq{k} \triangleq q_{m,n}^{\A}\br{t_k}\left|_{t_k = kT}\right.$ and $q^{\U}_{m,n}\sq{k} \triangleq q_{m,n}^{\U}\br{t_k}\left|_{t_k = kT}\right.$.

The one-frame Lyapunov drift and \mbox{drift-plus-penalty} functions \cite{Neelybook} are, respectively,
\begin{equation}\label{eqa:18}
\Delta_{\X}\sq{k} \triangleq \mathds{E}_{\X} \cb{ L\sq{k+1} -  L\sq{k} }
\end{equation}
and
\begin{equation}\label{eqa:19}
\Delta_{\X}\sq{k} + V  \sumtk \summ \mathds{E}_{\X}\cb{ G_m^{\sg}\br{t_k}}
\end{equation}
where $V > 0$ is a positive control parameter.

We obtain the upper bound of the  one-frame Lyapunov drift-plus-penalty function in \eqref{eqa:19} as
\begin{align}
    &  \Delta_{\X}\sq{k} + V \sumtk\summ \mathds{E}_{\X}\cb{ G_m^{\sg}\br{t_k} } \label{eqa:20}\\
\le & T\Psi +  V \sumtk\summ \mathds{E}_{\X}\cb{ G_m^{\sg}\br{t_k} } \nonumber\\
&+ \sumtk \br{\E^{\tt}_{\X}\cb{\v\br{t_k} - \r\br{t_k}}\q^{\A}\sq{k}} +  \sumtk \br{\E^{\tt}_{\X}\cb{\r\br{t_k} - \s\br{t_k}}\q^{\U}\sq{k}} \nonumber
\end{align}
where $\Psi \triangleq \frac{\br{s^{\max}}^2 + 2\br{r^{\max}}^2 + \br{\nu^{\max}}^2}{2}\sum\nolimits_{m=1}^M N_m$.
Please see Appendix \ref{apdx:01} for a detailed proof of the inequality \eqref{eqa:20}.

Minimizing the right-hand side (RHS) of \eqref{eqa:20} under the constraints in \eqref{eqa:06} and \eqref{eqa:12}--\eqref{eqa:14} gives us a feasible solution to the LTGEE minimization problem \eqref{eqa:16}.
Due to the constraints in \eqref{eqa:06} and \eqref{eqa:12}--\eqref{eqa:14}, the grid-energy expenditure is bounded by
\begin{equation}\label{eqa:21}
\abs{\sum\limits_{m=1}^M \mathds{E}_{\cal X}\cb{ G_m^{\sg}\br{t_k}}} \le \bar G
\end{equation}
where $\bar G$ is the bound of the grid-energy expenditure.

The properties on the obtained feasible solution is discussed in Proposition \ref{pr:01}.

\begin{proposition}[Asymptotic Optimality]\label{pr:01}
{\color{black}
Suppose that the initial queue backlogs $\q^{\A}\sq{0}$ and $\q^{\U}\sq{0}$ are fixed, and the resource allocation variables in $\Y$ satisfy the conditions
\begin{equation}\label{eqa:22}
\bar\v + \epsilon \bm 1 \le \E_{\X}\cb{\r\br{t_k}} \le \bar\s - \epsilon \bm 1, \forall t_k, k
\end{equation}
where $\epsilon$ is a small positive constant.
The NRE arrival rates are independent and stationary over the frames.
The channel-coefficient vectors and traffic arrival rates are independent and stationary over the slots.

When the above assumptions are satisfied, the minimizer to the RHS of \eqref{eqa:20} under the constraints in \eqref{eqa:06} and \eqref{eqa:12}--\eqref{eqa:14} asymptotically achieves the optimal grid-energy expenditure $G^*$ as
\begin{equation}\label{eqa:23}
G^* \le \frac{1}{KT}\sumk \sumtk \summ \mathds{E}_{\cal X}\cb{ G_m^{\sg}\br{t_k}} \le G^* + \frac{\Psi}{V}
\end{equation}
when the control parameter $V$ approaches infinity.
}

Moreover, the queue backlogs satisfy
\begin{equation}\label{eqa:24}
\limsup\limits_{K\rightarrow \infty}\frac{1}{K}\sumk \bm \E_{\X}\cb{q_{m,n}^{\A}\sq{k} + q_{m,n}^{\U}\sq{k}} \le \frac{\Psi + 2V\bar G }{\epsilon}
\end{equation}
such that the constraints in \eqref{eqa:15} are satisfied.
\end{proposition}
\begin{IEEEproof}
See Appendix \ref{apdx:02}.
\end{IEEEproof}

Based on Proposition \ref{pr:01}, we conclude that the set of minimizers $\Y^*$ to the RHS of \eqref{eqa:20} under the constraints in \eqref{eqa:06} and \eqref{eqa:12}--\eqref{eqa:14} is a feasible solution to the LTGEE minimization problem \eqref{eqa:16}.
Based on \eqref{eqa:23}, we observe that gap between the optimal grid-energy expenditure and obtained grid-energy expenditure by $\Y^*$ decreases with the control parameter as ${\cal O}\br{\frac{1}{V}}$.
Here, ${\cal O}\br{\frac{1}{V}}$ is a polynomial of $\frac{1}{V}$.
Based on Little's law and \eqref{eqa:24}, we observe that the end-to-end delay of UEs is a linearly increasing function of the control parameter $V$ as ${\cal O}\br{V}$. 
When the control parameter $V$ approaches infinity, the end-to-end delay of UEs increases to infinity.
Hence, we conclude that the grid-energy expenditure can be traded for the ene-to-end delay of UEs by tuning the control parameter.
Besides, the proposed analysis method in Appendix \ref{apdx:02} also explicitly defines the stable region of SGPCN as shown in \eqref{eqa:22}.



\section{Two-Scale UE Scheduling, Beamforming and Energy Exchanging}
Based on  Proposition \ref{pr:01}, we observe the elegance of a minimizer $\Y^*$ to the RHS of \eqref{eqa:20} under the constraints in \eqref{eqa:06} and \eqref{eqa:12}--\eqref{eqa:14}.
In this section, we propose a practical two-scale algorithm that jointly designs the scheduled UE indicators $\{a_{m,n}\sq{k}\}_{\forall m, n, k}$ in each frame and the beamforming vectors and exchanged NRE variables $\{\bm w_{m,n}\br{t_k}, \delta_m^l\br{t_k}\}_{\forall l, m, n, t_k, k}$ in each slot for the SGPCN.

\subsection{Optimal Scheduled UE Indicator in Each Frame}\label{nonpositive}
After some algebraic manipulations on the RHS of \eqref{eqa:20}, we obtain the term related to $\cb{ r_{m,n}\br{t_k}}_{\forall m,n,k,t_k}$ as
\begin{equation}\label{eqa:25}
\summ\sumn \br{ q_{m,n}^{\U}\sq{k} - q_{m,n}^{\A}\sq{k} } \mathds{E}_{\cal X}\cb{ \sumtk r_{m,n}\br{t_k} }.
\end{equation}
The term related to the grid-energy expenditure in the RHS of \eqref{eqa:20} is denoted by
\begin{equation}\label{eqa:26}
V\summ \mathds{E}_{\X}\cb{ \sumtk G_m^{\sg}\br{t_k} }.
\end{equation}

We observe that the terms $r_{m,n}\br{t_k}$ and $G_m^{\sg}\br{t_k}$ are coupled via $a_{m,n}\sq{k}$.
In order to minimize the RHS of \eqref{eqa:20}, we obtain the \emph{optimal} scheduled UE indicator as
\begin{equation}\label{eqa:27}
a_{m,n}^*\sq{k} = \left\{ \begin{array}{l}
0, q_{m,n}^{\U}\sq{k} - q_{m,n}^{\A}\sq{k} \ge 0 \mbox{ or } q_{m,n}^{\A}\sq{k} = 0 \\
1, \mbox{otherwise}.
\end{array} \right.
\end{equation}

{\color{black}
The optimality of \eqref{eqa:27} is proved in Appendix \ref{apdx:03}. }

\subsection{Optimal Beamforming and NRE Exchanging in Each Slot}
The NRE arrival rates $\cb{E_m^{\hav}\sq{k}}_{\forall m}$ remain constant during the $k$th frame, and  channel-coefficient vectors $\cb{\h_{m,n}\br{t_k}}_{\forall m, n}$ are independent over different slots in the $k$th frame.
Based on the principle of opportunistically minimizing an expectation \cite{Neelybook}, the minimizer to the RHS of \eqref{eqa:20} under the constraints in \eqref{eqa:06} and \eqref{eqa:12}--\eqref{eqa:14} can be obtained by solving a per-slot optimization problem as
\begin{equation}\label{eqa:28}
\begin{split}
 & \opt\br{t_k} = \\
 \min\limits_{\Z\br{t_k}} &\; \summ\sumn \br{q_{m,n}^{\U}\sq{k} - q_{m,n}^{\A}\sq{k} } { r_{m,n}\br{t_k}}  + V\summ { G_m^{\sg}\br{t_k} }  \\
\st &\; \eqref{eqa:06} \mbox{ and } \eqref{eqa:12}-\eqref{eqa:14}
\end{split}
\end{equation}
where $\Z\br{t_k} \triangleq \cb{\w_{m,n}\br{t_k}, \delta_m^l\br{t_k}}_{\forall m, n}$.

Solving the per-slot optimization problem \eqref{eqa:28} is challenging due to the non-convexity of  the rate-limit constraints in \eqref{eqa:12} and the proportional-rate constraints in \eqref{eqa:13}.

To handle the non-convex proportional-rate constraints in \eqref{eqa:13}, we introduce an auxiliary variable $\phi\br{t_k}$ such that $r_{m,n}\br{t_k} = q^{\A}_{m,n}\br{t_k}\phi\br{t_k}$.
Hence, we obtain the proportional-rate constraints in \eqref{eqa:13} as
\begin{align}
& \frac{\bm h^{\H}_{m,n}\br{t_k} \bm w_{m,n}\br{t_k}}{f_{m,n}\br{ \phi\br{t_k}} } \nonumber \\
& =  \sqrt{  I_{m,n}^{\intra}\br{t_k} +  I_{m,n}^{\inter}\br{t_k} + \sigma_{m,n}^2 }, n \in {\cal N}_{m}^{\act}\sq{k}, \forall m \label{eqa:29}\\
& \Im\br{\bm h^{\H}_{m,n}\br{t_k} \bm w_{m,n}\br{t_k}} = 0, n \in {\cal N}_{m}^{\act}\sq{k}, \forall m  \label{eqa:30}
\end{align}
where
\begin{equation}\label{eqa:31}
f_{m,n}\br{\phi\br{t_k}}  \triangleq \sqrt{ \exp\br{ q^{\A}_{m,n}\br{t_k}\phi\br{t_k} } - 1 }
\end{equation}
and $\Im\br{\cdot}$ denotes the imaginary part of a complex value.
Moreover, the range of $\phi\br{t_k}$ is set as $\sq{0, 1}$ to guarantee the constraints in \eqref{eqa:12}.

Replacing $r_{m,n}\br{t_k}$ by $q^{\A}_{m,n}\br{t_k}\phi\br{t_k}$ in the objective function \eqref{eqa:28}, we obtain
\begin{equation}\label{eqa:32}
\begin{split}
 & \obj_m\br{t_k} \\
=& V\br{a_b - a_s}\br{P_m^{\bs}\br{t_k} + E_m^{\lpe}\br{t_k} - \frac{1}{T}E_m^{\hav}\sq{k}}^+  \\
&+ Va_s\br{P_m^{\bs}\br{t_k} + E_m^{\lpe}\br{t_k} - \frac{1}{T}E_m^{\hav}\sq{k}} \\
&+ \sum\limits_{n \in \N_m^{\act}\sq{k}} \br{q_{m,n}^{\U}\sq{k} - q_{m,n}^{\A}\sq{k} } q_{m,n}^{\A}\br{t_k}\phi\br{t_k}
\end{split}
\end{equation}
where $P_m^{\bs}\br{t_k}$ is defined in \eqref{eqa:05}, and $E_m^{\lpe}\br{t_k}$ is defined in \eqref{eqa:07}.

Relaxing the constraints in \eqref{eqa:29}, we obtain a convex optimization problem as
\begin{subequations}\label{eqa:33}
\begin{align}
& \overline{\opt}\br{t_k} = \nonumber\\
\min\limits_{\Z\br{t_k}} &\; \summ \obj_m\br{t_k} \label{eqa:33a} \\
\st &\; \delta_{m}^{l}\br{t_k} + \delta_{l}^{m}\br{t_k} = 0, l \in \N_m, \forall m \label{eqa:33b}\\
&\; \frac{\bm h^{\H}_{m,n}\br{t_k} \bm w_{m,n}\br{t_k}}{f_{m,n}\br{ \phi\br{t_k}} }  \label{eqa:33c}\\
& \ge  \sqrt{  I_{m,n}^{\intra}\br{t_k} +  I_{m,n}^{\inter}\br{t_k} + \sigma_{m,n}^2 }, n \in {\cal N}_{m}^{\act}\sq{k}, \forall m \nonumber\\
&\; \Im\br{\bm h^{\H}_{m,n}\br{t_k} \bm w_{m,n}\br{t_k}} = 0, n \in {\N}_{m}^{\act}\sq{k}, \forall m \label{eqa:33d}\\
&\;   \sum\limits_{n \in {\cal N}_{m}^{\act}\sq{k}}  \left\|\bm w_{m,n}\br{t_k}\right\|_{\F}^2 \le P_m^{\max}, \forall m. \label{eqa:33e}
\end{align}
\end{subequations}

At each slot, the constraints in \eqref{eqa:33b}--\eqref{eqa:33e} constitute a convex hull of the constraints in \eqref{eqa:06} and \eqref{eqa:12}--\eqref{eqa:14} when the values of $\phi\br{t_k}$ and $\cb{a_{m,n}\sq{k}}_{\forall m,n}$ are fixed.
Therefore, we conclude $\overline{\opt}\br{t_k} \le \opt\br{t_k}$.
Based on the arguments in \cite{DongJan.2019}, we demonstrate that the optimal $\Z^*\br{t_k}$ make the constraints in \eqref{eqa:33c} active.
In other words, we obtain $\overline{\opt}\br{t_k} = \opt\br{t_k}$.
See Appendix \ref{apdx:04} for a detailed proof of the activeness of constraints in \eqref{eqa:33c}.
{\color{black}
Motivated by Proposition 2 of \cite{DongJan.2019}, the optimal $\phi^*\br{t_k}$ can be obtained via a one-dimensional search method.
Therefore, the optimization problem \eqref{eqa:28} can be optimally solved. }

Based on \eqref{eqa:27}, the scheduled UE indicators are updated at the start of each frame.
Performing a one-dimensional search and solving the optimization problem \eqref{eqa:33}, the beamforming vectors and exchanged NRE variables are updated at the start of each slot.
Therefore, we summarize the TSUBE algorithm in Algorithm \ref{alg:01}.

\begin{algorithm}[ht]\small
  \centering
  \caption{TSUBE Algorithm}\label{alg:01}
  \begin{algorithmic}[1]
  \State \textbf{Inputs:} Harvested NRE $\cb{E_m^{\hav}\sq{k}}_{\forall m}$,
  channel-coefficient vectors $\cb{\h_{m,n}\br{t_k}}_{\forall m,n}$,
  traffic arrival rate $\cb{\nu_{m,n}\br{t_k}}_{\forall m,n}$,
  backlogs of access queues $\cb{q_{m,n}^{\A}\sq{k}}_{\forall m,n}$ and
  processing queues ${q_{m,n}^{\U}\sq{k}}_{\forall m,n}$
  \State At the start of the $k$th frame, the CN estimates the harvested NRE as $\cb{E_m^{\hav}\sq{k}}_{\forall m}$
  \State At the start of the $k$th frame, the CN updates the scheduled UE indicators $\cb{a_{m,n}\sq{k}}_{\forall m, n}$ via \eqref{eqa:27}, backlog of access queues $\cb{q_{m,n}^{\A}\sq{k}}_{\forall m,n}$ and backlogs of processing queues ${q_{m,n}^{\U}\sq{k}}_{\forall m,n}$
  \Repeat \label{alg1:line1}
  \State At the start of the $t_k$th slot, the CN estimates the channel-coefficient vectors $\cb{\bm h_{m,n}\br{t_k}}_{\forall m, n}$
  \State Based on  $\cb{\bm h_{m,n}\br{t_k}}_{\forall m, n}$ and $\cb{a_{m,n}\sq{k}}_{\forall m,n}$, the CN solves the optimization problem \eqref{eqa:33} via CVX \cite{Grant2014}
  \State At the start of the $t_k$th slot, the CN performs one dimensional search for the optimal $\phi^*\br{t_k}$ \label{alg1:line3}
  \Until{The optimal $\phi^*\br{t_k}$ is obtained}\label{alg1:line2}
  \State At the end of the $t_k$th slot, the CN updates the access queues and processing queues according to \eqref{eqa:09} and \eqref{eqa:10}
  \State \textbf{Outputs:} Scheduled UE indicators $\cb{\alpha_{m,n}\sq{k}}_{\forall m,n}$,
  beamforming vectors $\cb{\w_{m,n}\br{t_k}}_{\forall m,n}$ and
  exchanged NRE variables $\cb{\delta_m^l\br{t_k}}_{m,l}$
  \end{algorithmic}
\end{algorithm}

{\color{black}
\emph{Complexity Analysis:}
For brevity, we assume that $N = N_m$, $m = 1, \ldots, M$.
The number of UEs in the SGPCN is $MN$.
The complexity of scheduling UEs via \eqref{eqa:27} is calculated as $MN$ at the start of each frame.

The major complexity of the TSUBE Algorithm per slot lies in the iteration loop in lines \ref{alg1:line1}--\ref{alg1:line2}.
Hereinafter, we focus on analyzing the computational complexity of the iteration loop.
Moreover, the computational complexity of the iteration loop in lines \ref{alg1:line1}--\ref{alg1:line2} comes from solving \eqref{eqa:33} via the interior-point method and the one-dimensional search.
Hence, we evaluate the worst-case computational complexity of solving \eqref{eqa:33} via the interior-point method and multiply it by the number of points in the one-dimensional search to obtain the computational complexity of the iteration loop in lines \ref{alg1:line1}--\ref{alg1:line2}.
We observe that the optimization problem \eqref{eqa:33} is second-order conic programming.
In the optimization problem \eqref{eqa:33}, the number of second-order cones with dimension $LMN$ is $M$, and the number of second-order cones having dimension $LN$ is $M$.
The number of linear constraints is $m_1 = MN + \frac{1}{2}\sum\nolimits_{m=1}^M\abs{\N_m}$.
The number of variables is $m_2 = LMN + \frac{1}{2}\sum\nolimits_{m=1}^M\abs{\N_m}$ in the optimization problem \eqref{eqa:33}.
According to \cite[Lecture 6]{Ben-Tal2001}, an $\epsilon$-accurate solution to \eqref{eqa:33} requires $n_1 = \OO{\log\epsilon^{-1}\sqrt{MN + 2M + \frac{1}{2}\sum\nolimits_{m=1}^M\abs{\N_m}}}$ iterations, and the computational complexity per iteration is $\OO{\br{m_2+1}m_2m_1 + m_2m_3 + m_2^3}$ where $m_3 = MN\br{M+1}$.
Therefore, the computational complexity of the iteration loop in lines \ref{alg1:line1}--\ref{alg1:line2} is $\OO{\Xi n_1m_2\br{\br{m_2+1}m_1 + m_3 + m_2^2}}$ where $\Xi$ is the number of points for a one-dimensional search.
}

\section{Numerical Results}
In this section, we present simulation results to evaluate the proposed TSUBE algorithm.
The pathloss of the $\br{m,n}$th access link is calculated as
\begin{equation}\label{eqa:34}
\omega_{m,n} = 17.3 + 38.3\log_{10}d_{m,n} + 24.9\log_{10}f_c \mbox{  dB}
\end{equation}
where $d_{m,n}$ is the link distance of the $\br{m,n}$th access link, and carrier frequency $f_c = 2.1$ GHz.

We consider a two-BST SGPCN, where each BST is associated with three UEs and is equipped with six antennas.
The SGPCN operates in two time scales, where each frame consists of five slots.
The inter-BST distance is set as $400$ meters.
The UEs are deployed at the middle point between the two BSTs such that the worst-case interference is considered.
The power of AWGN noise is set as $1\times 10^{-10.7}$ mW.
The power amplifier efficiency, maximum transmit power, baseband processing power of BSTs are, respectively, set as $\eta = 0.8$, $P_m^{\max} = 400$ mW and $P_m^{\sp} = 100$ mW.
The efficiency of local power lines are set as $\beta_m^l = 0.8$.
Unless otherwise specified, the purchasing and selling prices of a unit energy is set as $\alpha_b = 1.6 \times 10^{-9}$ cents/slot/mW and $\alpha_s = 0.6 \times 10^{-9}$ cents/slot/mW.
The average arrival rate $\bar\nu_{m,n}$ and constant  processing rate $\tilde s_{m,n}$ are, respectively, set as $2.1$ nats/slot/Hz and $8$ nats/slot/Hz.
The average NRE arrival rates of the first BST and the second BST are set as $300$ mW/slot and $200$ mW/slot.
The grid-energy expenditure is annualized with the duration of a slot as $1$ ms and number of BSTs as $1\times 10^3$.
{\color{black}
The value of control parameter $V$ is empirically tuned to demonstrate tradeoff between the end-to-end delay and grid-energy expenditure.}

{\color{black}
We consider two benchmark schemes, namely WOLPE \cite{Dongtobe} and zero-forcing beamforming (ZFBF) algorithm.
Denote ${\bm H}_{m,n}\br{t_k} \triangleq [\h_{m,i}\br{t_k}]_{i = 1:N, i\neq n}$ where $N = \sum\nolimits_{m = 1}^M N_m$.
Performing the singular-value decomposition on ${\bm H}_{m,n}^{\H}\br{t_k} \in \C^{\br{N-1}\times L}$, we obtain the null-space basis matrix of ${\bm H}_{m,n}\br{t_k}$ as $\XI_{m,n}\br{t_k} \in \C^{L\times \br{L-N+1}}$ with $\XI_{m,n}^{\H}\br{t_k}\XI_{m,n}\br{t_k} = \bm I$.
Moreover, we align the $\br{m,n}$th ZFBF vector to the channel-coefficient vectors $\h_{m,n}\br{t_k}$ as
\begin{equation}
\w_{m,n}\br{t_k} = \sqrt{p_{m,n}\br{t_k}}\frac{\XI_{m,n}\br{t_k}\XI^{\H}_{m,n}\br{t_k}\h_{m,n}\br{t_k}}
{\norm{\XI^{\H}_{m,n}\br{t_k}\h_{m,n}\br{t_k}}}
\end{equation}
where $p_{m,n}\br{t_k}$ is the transmit power for the $\br{m,n}$th UE at the $t_k$th slot.
The received signal-to-noise ratio of the $\br{m,n}$th UE at the $t_k$th slot is
\begin{equation}
\mbox{SNR}_{m,n}\br{t_k} = p_{m,n}\br{t_k}\frac{\norm{\XI^{\H}_{m,n}\br{t_k}\h_{m,n}\br{t_k}}^2}{\sigma_{m,n}^2}.
\end{equation}

When the ZFBF vectors are used, problem \eqref{eqa:33} is reduced to an optimization problem having scalar variables.
Hence, the ZFBF algorithm has a lower computational complexity than the TSUBE algorithm.
}

\begin{figure}[ht]
\centering
  \includegraphics[width= 3.3 in]{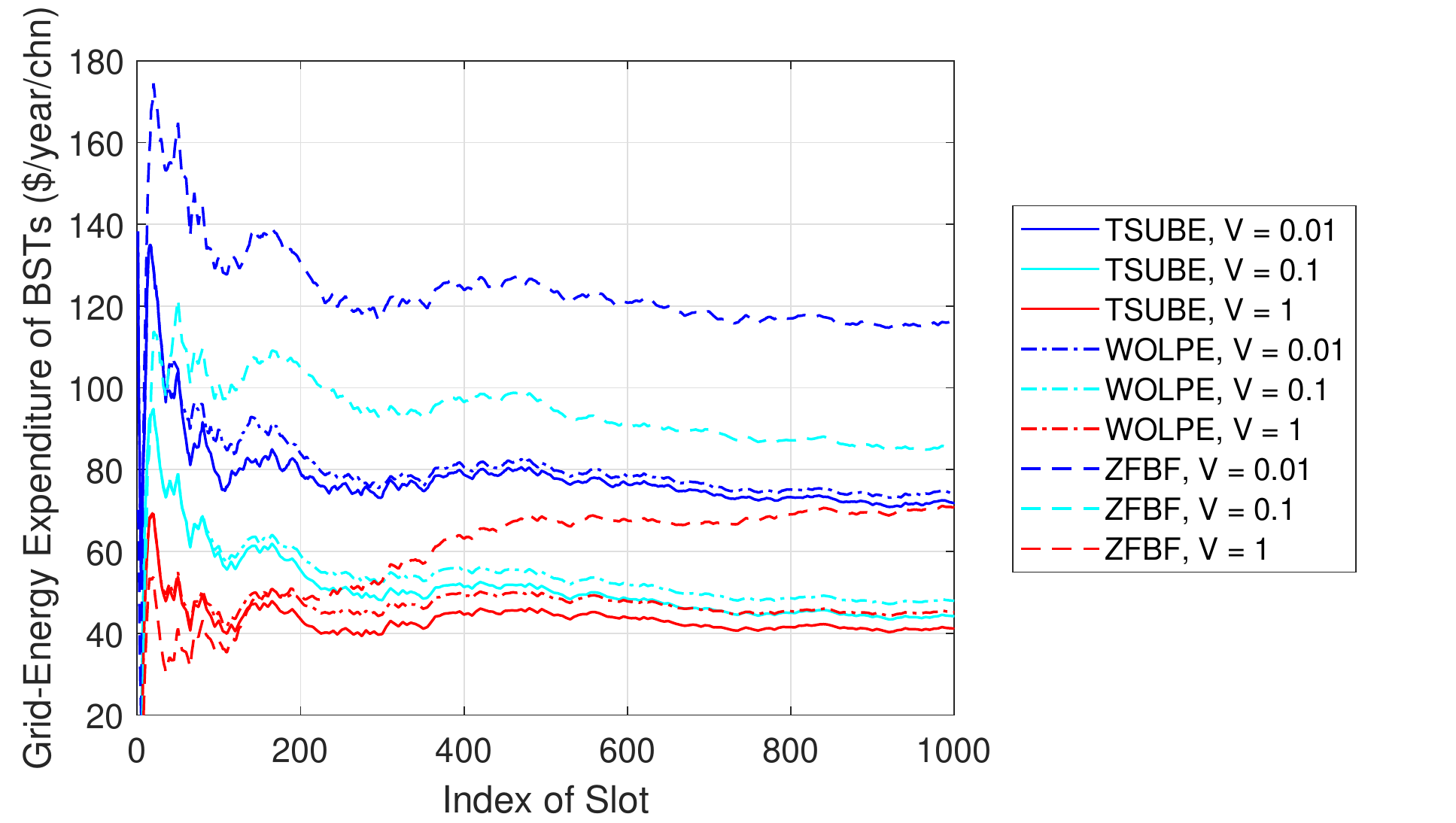}
  \caption{The moving-average grid-energy expenditure with window size = $10$.}\label{fg:002}
  \includegraphics[width= 3.3 in]{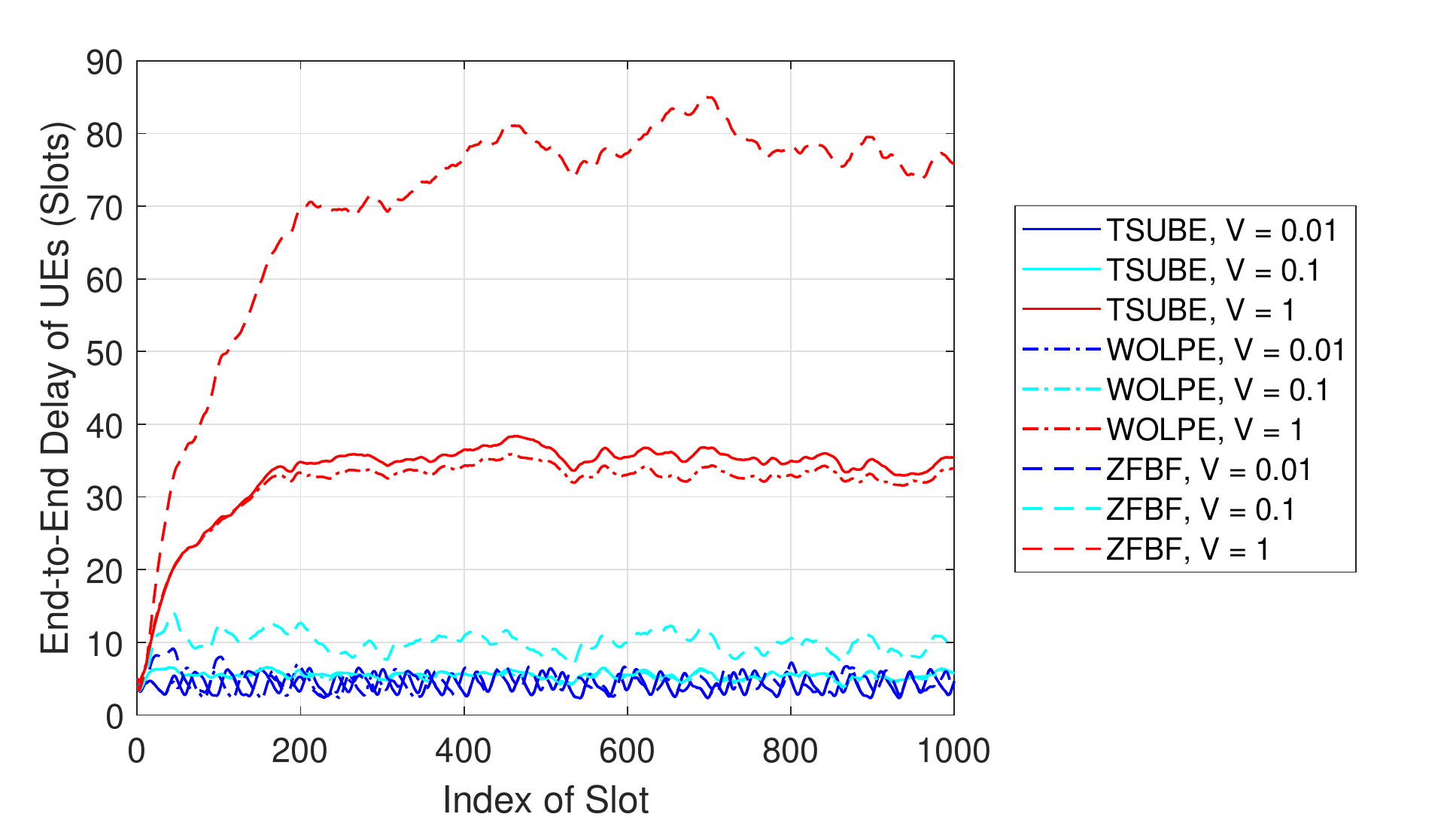}
  \caption{The moving-average end-to-end delay with window size = $10$.}\label{fg:003}
\end{figure}

{\color{black}
Figures \ref{fg:002} and \ref{fg:003} show the moving-average annualized grid-energy expenditure and moving-average end-to-end delay of UEs when the moving-average window is set as $10$.
We observe that the moving-average grid-energy expenditures of the proposed TSUBE algorithm, the WOLPE algorithm and the ZFBF algorithm converge within $1,000$ slots.
The moving-average end-to-end delay of UEs becomes stable after $400$ slots.
Note that the end-to-end delay is calculated according to Little's law for the two cascading queues.
When the control parameter $V$ is set as $0.01$, $0.1$ and $1$, the grid-energy expenditures of the proposed TSUBE algorithm are, respectively, $3.15\%$, $7.85\%$ and $8.85\%$ lower  than that of the WOLPE algorithm, and $37.67\%$, $48.12\%$ and $41.82\%$ lower than that of the ZFBF algorithm.
This observation is due to the facts that: 1) the proposed TSUBE algorithm intelligently makes decisions on whether to purchase grid-energy or exchange NRE to avoid redundant grid-energy transactions; and 2) the WOLPE algorithm introduces redundant purchasing/selling of grid-energy when the SGPCN has deficit/surplus NRE;
3) the ZFBF algorithm prefers to mitigate interference.

\begin{figure}[ht]
\centering
  \subfigure[Grid-energy expenditure v.s. control parameter]{\includegraphics[width=3.3 in]{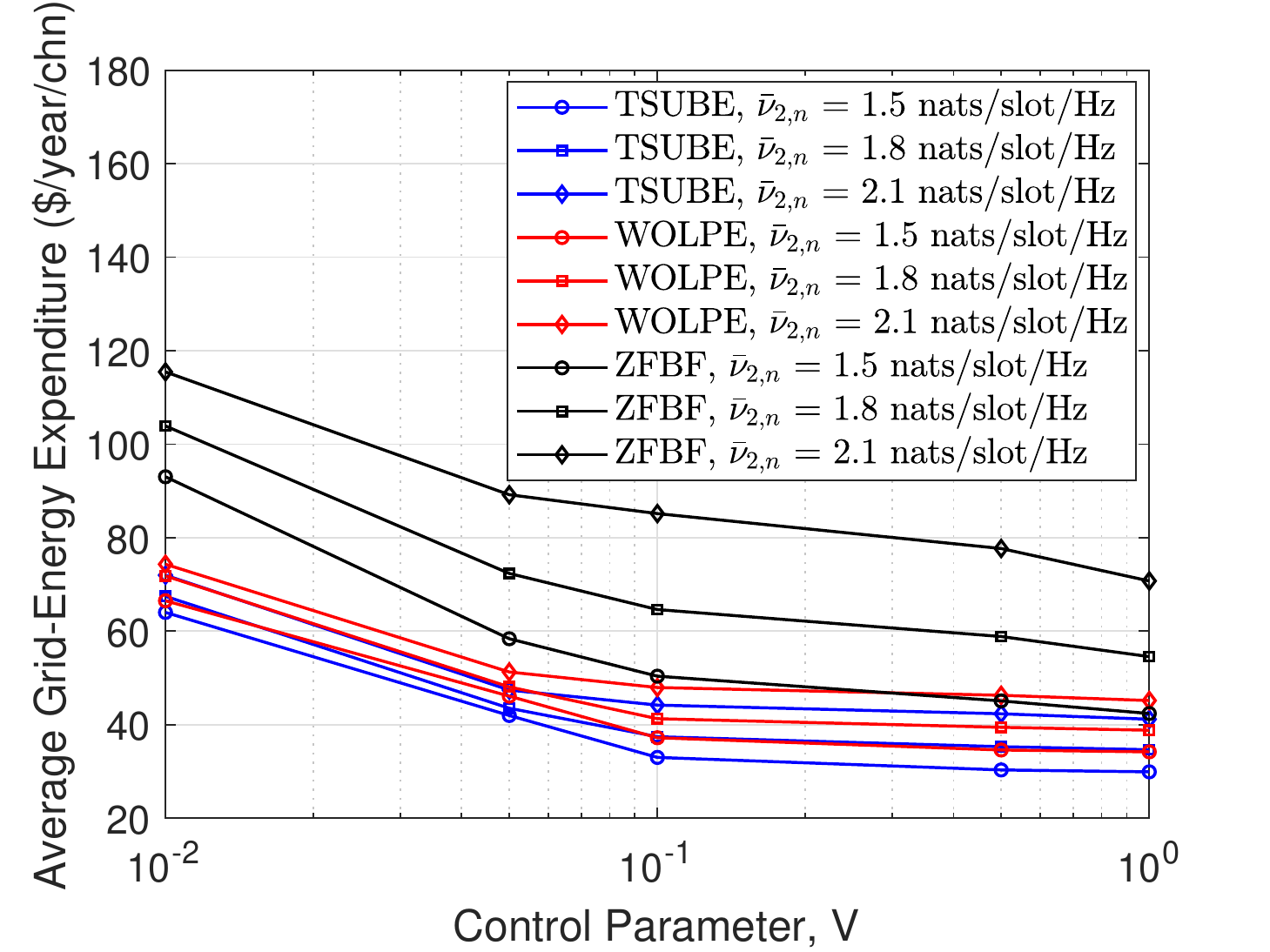}\label{fg:004a}}
  \subfigure[End-to-end delay v.s. control parameter]{\includegraphics[width=3.3 in]{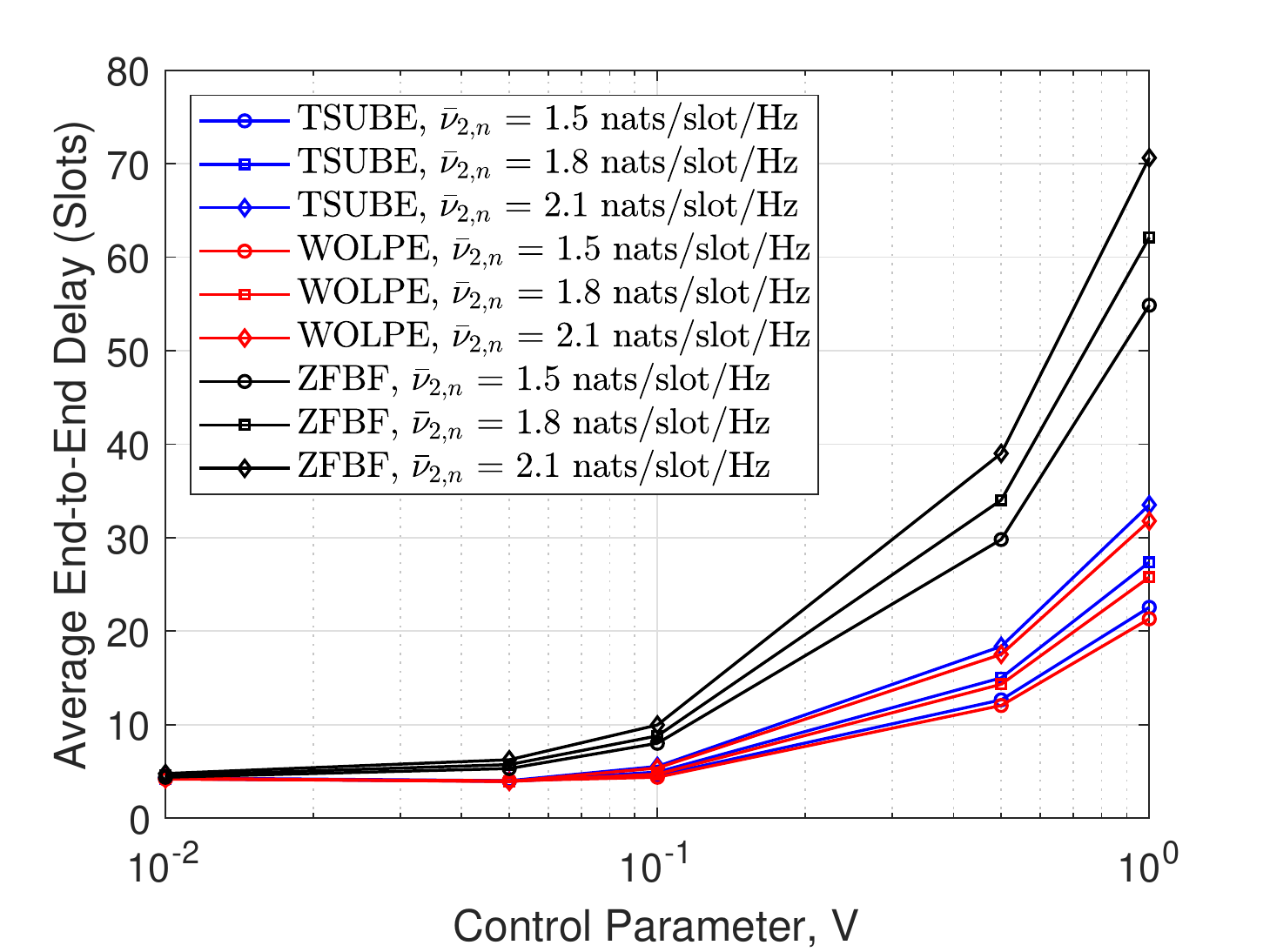}\label{fg:004b}}
  \caption{The tradeoff between the average grid-energy expenditure and average end-to-end delay of UEs.}\label{fg:004}
\end{figure}

Figure \ref{fg:004} reveals the tradeoff between the average grid-energy expenditure and the end-to-end delay of UEs under different average arrival rates of UEs in the second BST (i.e., $\bar\nu_{2, n}$).
We observe that increasing the control parameter induces a decreasing grid-energy expenditure (as shown in Fig. \ref{fg:004a}) and an increasing end-to-end delay of UEs (as shown in Fig. \ref{fg:004b}).
Therefore, the proposed TSUBE algorithm, WOLPE algorithm and ZFBF algorithm provide the operator with flexibility in controlling the grid-energy expenditure while maintaining a satisfactory level of communication QoS.
Moreover, we also observe that the proposed TSUBE algorithm outperforms the WOLPE algorithm and ZFBF algorithm in terms of the grid-energy expenditure.
For example, when $V = 0.1$ and $\bar\nu_{2, n} = 1.5$ nats/slot/Hz, the TSUBE algorithm achieves $11.32\%$ lower grid-energy expenditure than the WOLPE algorithm by sacrificing $3.86\%$ the end-to-end delay of UEs.
When $V = 1$ and $\bar\nu_{2, n} = 1.5$ nats/slot/Hz, the TSUBE algorithm achieves $12.51\%$ lower grid-energy expenditure than the WOLPE algorithm by sacrificing $5.45\%$ the end-to-end delay of UEs.
Moreover, the TSUBE algorithm outperforms the ZFBF algorithm in terms of grid-energy expenditure and end-to-end delay.
For example, when $V = 0.1$ and $\bar\nu_{2, n} = 1.8$ nats/slot/Hz, the TSUBE algorithm achieves $35.08\%$ lower grid-energy expenditure and $7.41\%$ lower end-to-end delay than the ZFBF algorithm.
This observation is due to the fact that the local power exchanging introduces a new dimension of freedom to reduce the grid-energy expenditure when the SGPCN has a more stringent energy demand.
When more NRE is traded to reduce the grid-energy expenditure, the end-to-end delay of UEs increases.

\begin{figure}[ht]
\centering
  \includegraphics[width= 3.6 in]{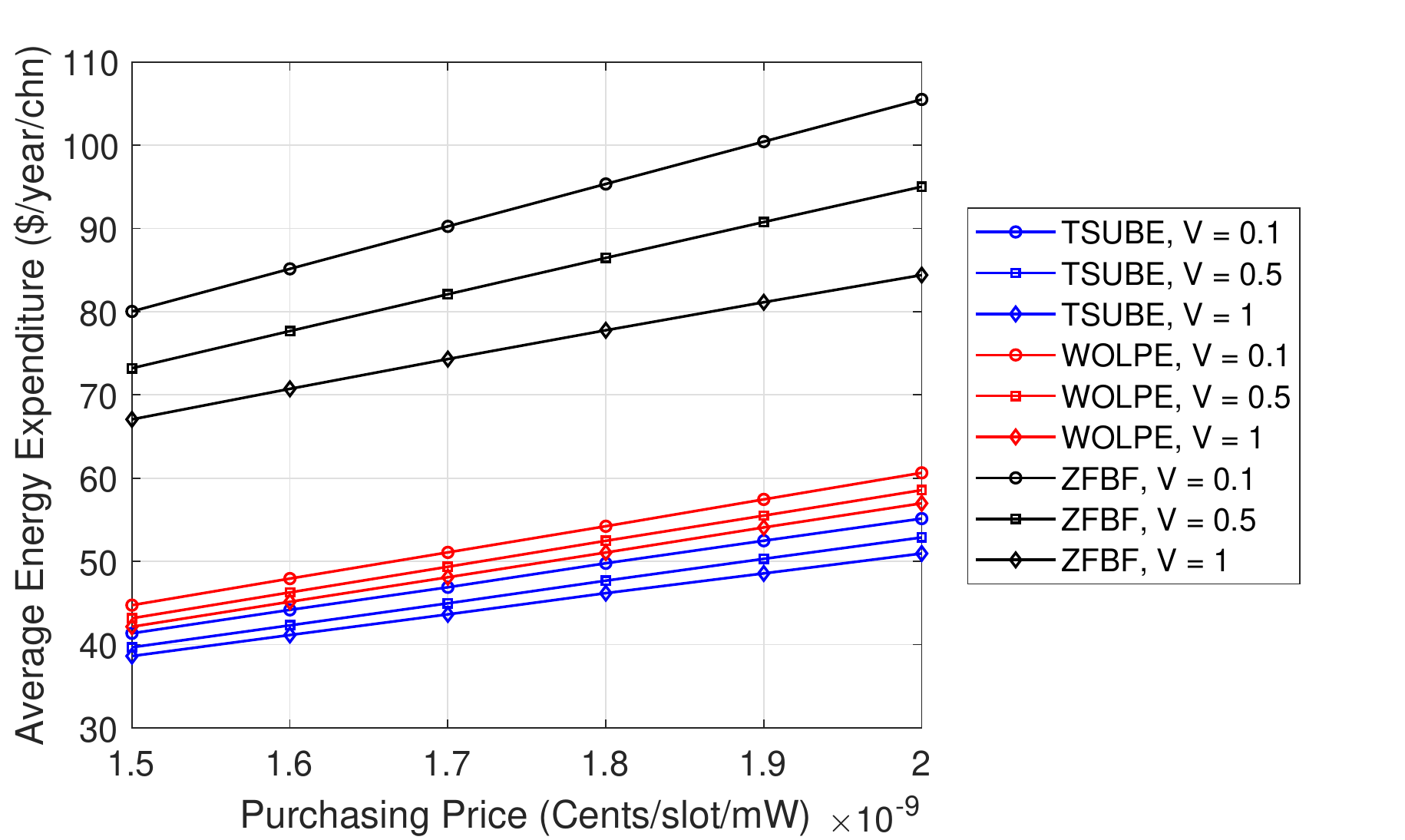}
  \caption{The average grid-energy expenditure versus the purchasing price $\alpha_b$.}\label{fg:007}
\end{figure}

Figure \ref{fg:007} shows that grid-energy expenditure increases with the purchasing price of a unit energy under various control parameters.
More specifically, the gap of grid-energy expenditure between the proposed TSUBE algorithm and WOLPE algorithm increases with the purchasing price $\alpha_b$.
The reason is as follows.
A higher purchasing price $\alpha_b$ motivates the BSTs to exchange NRE via the local power line such that the grid-energy expenditure of TSUBE algorithm increases slower than that of the WOLPE algorithm.
Compared with the WOLPE algorithm, the proposed TSUBE algorithm can reduce the grid-energy expenditure by $9.07\%$, $9.71\%$ and $10.58\%$ when the control parameters are respectively set as $0.1$, $0.5$ and $1$.
In other words, a higher control parameter induces a more effective grid-energy expenditure reduction of TSUBE algorithm than the WOLPE algorithm.
Besides, we also observe that grid-energy expenditure of TSUBE algorithm is lower than that of the ZFBF algorithm.
This is due to the fact that the ZFBF algorithm requires the BSTs to consume more grid-energy than the TSUBE algorithm to guarantee the stability of SGPCN.

\begin{figure}[ht]
\centering
  \includegraphics[width= 3.3 in]{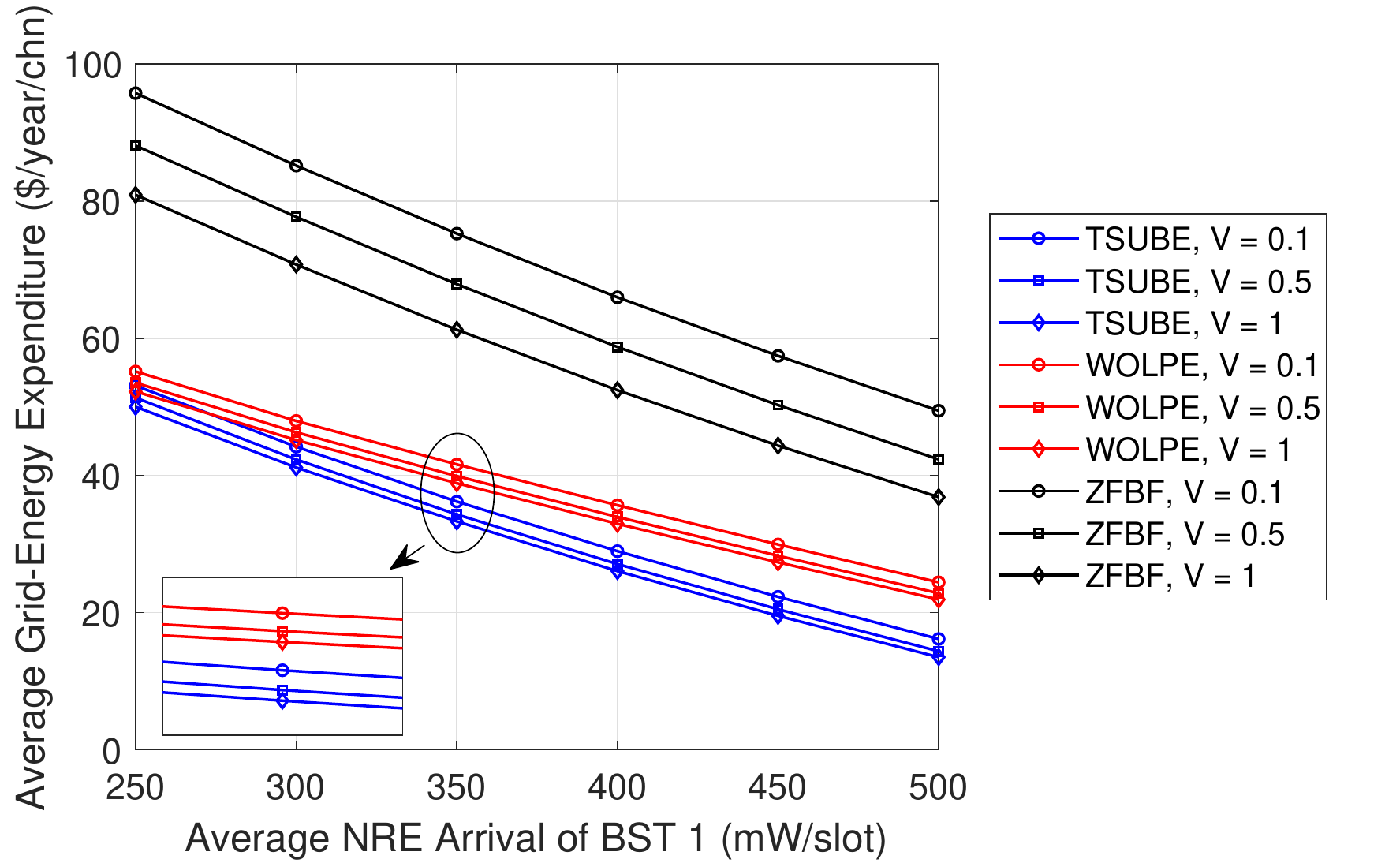}
  \caption{The average grid-energy expenditure versus the average NRE arrival rate of the first BST.}\label{fg:006}
\end{figure}

Figure \ref{fg:006} illustrates the grid-energy expenditure as a function of the average NRE arrival rate of the first BST under different control parameters.
Increasing the average NRE arrival rate of the first BST from $250$ mW/slot to $500$ mW/slot, we observe that the grid-energy expenditures of the proposed TSUBE algorithm, WOLPE algorithm and ZFBF algorithm decrease.
Moreover, by increasing the average NRE arrival rate, we also observe that the gaps of grid-energy expenditure between the proposed TSUBE algorithm and WOLPE algorithm increase from
$2.07$  \$/year/chn, $2.18$ \$/year/chn and $2.22$ \$/year/chn to
$8.25$ \$/year/chn, $8.45$ \$/year/chn and $8.40$ \$/year/chn  when the control parameters are respectively set as $V = 0.1$, $V = 0.5$ and $V = 1$.
These observations demonstrate that the proposed TSUBE algorithm outperforms the WOLPE algorithm.
Moreover, the proposed TSUBE algorithm can reduce the grid-energy expenditure by $72.99\%$ when the control parameter and average NRE arrival rate are $V = 1$ and $500$ mW/slot.
Since the NRE arrival rate of the second BST is $200$ mW/slot, we conclude that a more asymmetric NRE arrival rate induces a more frequent local energy exchange under symmetric data rate.
Therefore, the gaps of grid-energy expenditure between the proposed TSUBE algorithm and WOLPE algorithm can increase with the average NRE arrival rate of the first BST.
Figure \ref{fg:006} also shows that the proposed TSUBE algorithm outperforms the ZFBF algorithm when the average NRE arrival rate increases.
The gaps of grid-energy expenditure are as large as $42.64$ \$/year/chn, $36.75$ \$/year/chn and $30.89$ when the control parameters are respectively $0.1$, $0.5$ and $1$.
This observation indicates that the wireless operator can choose the ZFBF algorithm for low computational complexity at the expense of grid-energy expenditure.

\begin{figure}[ht]
\centering
\includegraphics[width= 3.6 in]{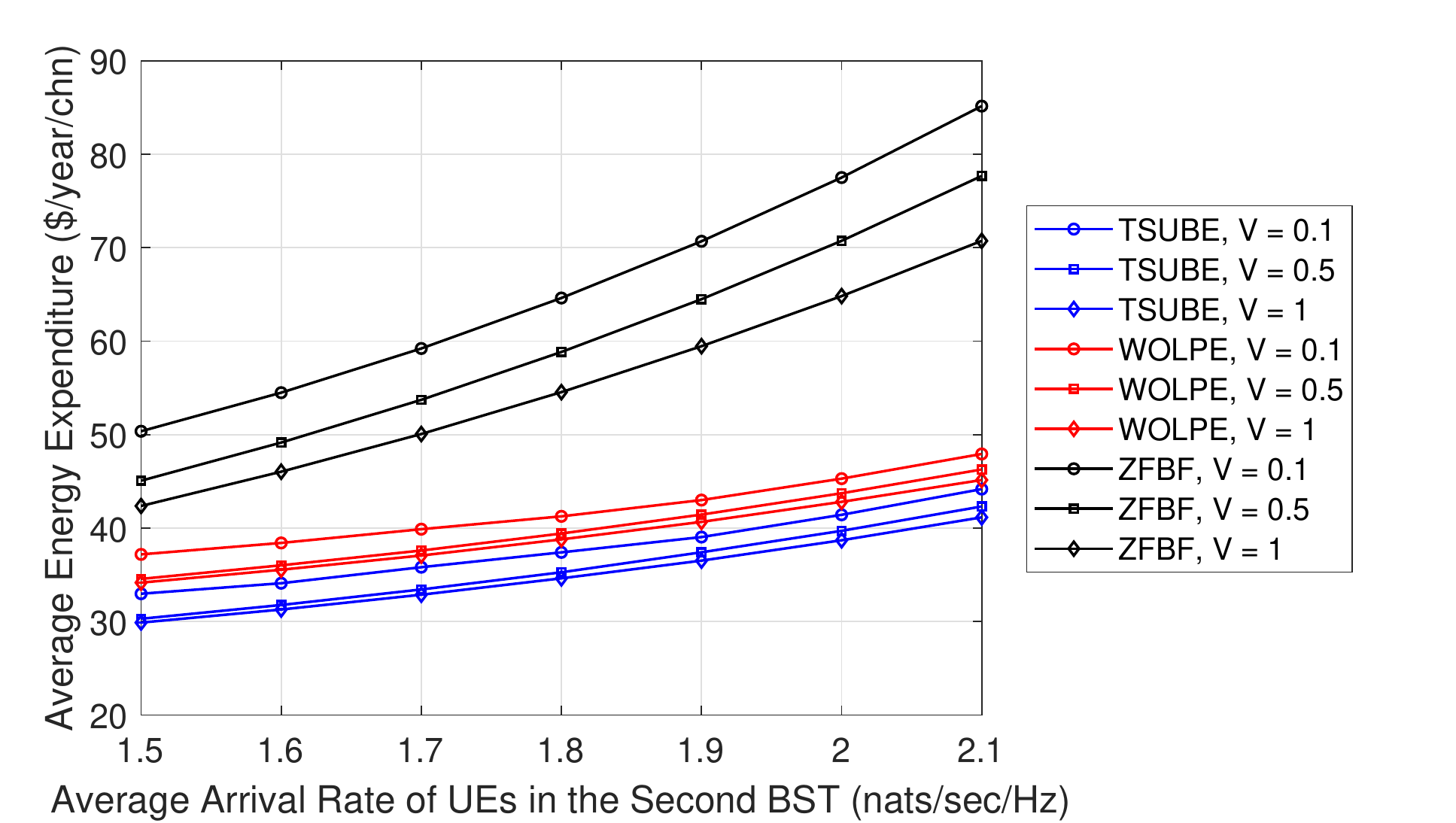}
\caption{The average grid-energy expenditure versus the average arrival rate $\bar\nu_{2, n}$.}\label{fg:005}
\end{figure}

Figure \ref{fg:005} shows that the grid-energy expenditure increases with the average arrival rate of UEs in the second BST under different control parameters.
A higher data arrival rate induces a higher grid-energy expenditure.
Therefore, we observe that the grid-energy expenditure increases with the average data arrival rate of the second BST in Fig. \ref{fg:005}.}

\section{Conclusions}
We have investigated the LTGEE minimization problem in  SGPCNs, and proposed a TSUBE algorithm for SGPCNs to  allocate jointly the scheduled UE indicators, beamforming vectors and exchanged NRE variables.
We have leveraged the Lyapunov optimization method to decouple the beamforming vectors design and scheduled UE indicators allocation.
Based on the proposed TSUBE algorithm, the scheduled UE indicators are optimally allocated at each frame in order to avoid redundant scheduling/unscheduling user equipments.
The beamforming vectors and exchanged NRE variables are optimally allocated to minimize the per-slot subproblems.
When the control parameter approaches infinity, the proposed TSUBE algorithm asymptotically achieves the optimal grid-energy expenditure.
The tradeoff between the grid-energy expenditure and the end-to-end delay of UEs has been theoretically established when three dimensions of resources (scheduled UE indicators, beamforming vectors and exchanged NRE variables) are jointly allocated.
Numerical results have been presented to demonstrate that the TSUBE algorithm outperforms the WOLPE algorithm and ZFBF algorithm in terms of grid-energy expenditure.
Therefore, the joint allocation of three-dimensional resources (scheduled UE indicators, beamforming vectors and exchanged NRE variables) helps to reduce grid-energy expenditure and yields a better tradeoff between grid-energy expenditure and UE data rates compared with the joint allocation of two-dimensional resources (scheduled UE indicators and beamforming vectors).

\appendices
\section{Proof of \eqref{eqa:20}}\label{apdx:01}
Taking the telescoping summation over $kT \le t_k < \br{k+1}T - 1$ for the $\br{m,n}$th access queue in \eqref{eqa:09}, we obtain the one-frame dynamic equation of the $\br{m,n}$th access queue as
\begin{equation}\label{eqa:apdx:01}
\begin{split}
&q_{m,n}^{\A}\sq{k+1} \\
=& q_{m,n}^{\A}\sq{k}  + \sumtk \nu_{m,n}\br{t_k} - \sumtk r_{m,n}\br{t_k}.
\end{split}
\end{equation}
Based on \eqref{eqa:apdx:01}, the one-frame drift of the $\br{m,n}$th access queue is upper-bounded as
\begin{equation}\label{eqa:apdx:02}
\begin{split}
& \frac{1}{2}\br{\br{ q_{m,n}^{\A}\sq{k+1} }^2 - \br{ q_{m,n}^{\A}\sq{k} }^2} \\
\le& \frac{\br{\nu^{\max}}^2 + \br{r^{\max}}^2}{2}T  \\
& + q_{m,n}^{\A}\sq{k}  \sumtk \br{\nu_{m,n}\br{t_k}  -  r_{m,n}\br{t_k}}.
\end{split}
\end{equation}
where the inequality holds due to the facts in \eqref{eqa:11}.

Following a similar argument, we obtain the upper-bound of the one-frame drift of the \mbox{$\br{m,n}$th} processing queue as
\begin{equation}\label{eqa:apdx:03}
\begin{split}
& \frac{1}{2}\br{\br{q_{m,n}^{\U}\sq{k+1}}^2 - \br{q_{m,n}^{\U}\sq{k}}^2} \\
\le&  \frac{\br{s^{\max}}^2 + \br{r^{\max}}^2}{2}T \\
& + q_{m,n}^{\U}\sq{k}\sumtk\br{  r_{m,n}\br{t_k} - s_{m,n}\br{t_k} }
\end{split}
\end{equation}
where the inequality holds due to the facts in \eqref{eqa:11}.

Based on \eqref{eqa:apdx:02} and \eqref{eqa:apdx:03}, we obtain the \mbox{upper-bound} of the one-frame Lyapunov drift-plus-penalty function in \eqref{eqa:19} as
\begin{equation}\label{eqa:apdx:05}
\begin{split}
    &  \Delta_{\X}\sq{k} + V \sumtk\summ \E_{\X}\cb{ G_m^{\sg}\br{t_k} } \\
\le & T\Psi + V \sumtk\summ \E_{\X}\cb{ G_m^{\sg}\br{t_k}} \\
&+ \sumtk \E^{\tt}_{\X}\cb{\v\br{t_k} - \r\br{t_k}}\q^{\A}\sq{k}  \\
& + \sumtk \E^{\tt}_{\X}\cb{\r\br{t_k} - \s\br{t_k}}\q^{\U}\sq{k}
\end{split}
\end{equation}
where $\Psi \triangleq \frac{\br{s^{\max}}^2 + 2\br{r^{\max}}^2 + \br{\nu^{\max}}^2}{2}\sum\nolimits_{m=1}^M N_m$.

Rearranging the RHS of \eqref{eqa:apdx:05}, we obtain \eqref{eqa:20}.

\section{Proof of Proposition \ref{pr:01}}\label{apdx:02}
Let $\Y^* = \cb{\Y^*\br{t_k}}_{\forall t_k, k}$ denote the set of optimal resource allocation variables, which minimize the RHS of \eqref{eqa:20} under the constraints in \eqref{eqa:06} and \eqref{eqa:12}--\eqref{eqa:14}.
Let $\tilde\Y = \cb{\tilde\Y\br{t_k}}_{\forall t_k, k}$ denote the set of feasible resource allocation variables such that $\bar\v + \epsilon\bm 1 \le \E_{\X}\cb{\r\br{\tilde\Y\br{t_k}} } \le \bar\s - \epsilon\bm 1$.

Substituting the set of optimal resource allocation variables $\Y^*$ into the RHS of \eqref{eqa:19}, we obtain
\begin{align}
& \Delta_{\X}\sq{k} + V \sumtk\summ \mathds{E}_{\X}\cb{ G_m^{\sg}\br{\Y^*\br{t_k}} } \label{eqa:apdx2:01a}\\
\le & T\Psi + V \sumtk\summ \mathds{E}_{\X}\cb{ G_m^{\sg}\br{\Y^*\br{t_k}} } \nonumber\\
& + \sumtk \E^{\tt}_{\X}\cb{\v\br{t_k} - \r\br{\Y^*\br{t_k}}}\q^{\A}\sq{k} \nonumber\\
& + \sumtk \E^{\tt}_{\X}\cb{\r\br{\Y^*\br{t_k}} - \s\br{t_k}}\q^{\U}\sq{k} \label{eqa:apdx2:01b}\\
\le & T\Psi + V \sumtk\summ \mathds{E}_{\X}\cb{ G_m^{\sg}\br{\tilde\Y\br{t_k}}} \nonumber\\
&+ \sumtk \E^{\tt}_{\X}\cb{\v\br{t_k} - \r\br{\tilde\Y\br{t_k}}}\q^{\A}\sq{k} \nonumber\\
& + \sumtk \E^{\tt}_{\X}\cb{\r\br{\tilde\Y\br{t_k}} - \s\br{t_k}}\q^{\U}\sq{k} \label{eqa:apdx2:01c} \\
\le & T\Psi + V \sumtk\summ \mathds{E}_{\X}\cb{ G_m^{\sg}\br{\tilde\Y\br{t_k}}} \nonumber\\
& - \epsilon T\bm 1^{\tt}\br{\q^{\A}\sq{k} + \q^{\U}\sq{k}} \label{eqa:apdx2:01d}
\end{align}
where the inequality \eqref{eqa:apdx2:01c} follows the fact that $\Y^*$ is the minimizer to the RHS of \eqref{eqa:apdx2:01b};
and the inequality \eqref{eqa:apdx2:01d} follows the fact $\bar\v + \epsilon\bm 1 \le \E_{\X}\cb{\r\br{\tilde\Y\br{t_k}}} \le \bar\s - \epsilon\bm 1$.

Rearranging \eqref{eqa:apdx2:01d} and taking iterated expectation over random sources $\X$, we obtain an upper bound of the one-frame Lyapunov drift function as
\begin{equation}\label{eqa:apdx2:02}
\begin{split}
 & \E_{\X}\cb{L\sq{k+1} - L\sq{k}} \\
\le &  T\Psi + 2 TV\bar G - \epsilon T\bm 1^{\tt}\E_{\X}\cb{\q^{\A}\sq{k} + \q^{\U}\sq{k}}
\end{split}
\end{equation}
based on the bounded grid-energy expenditure \eqref{eqa:21}.

Taking telescoping summation over $k = 0, 1, \ldots, K-1$ for \eqref{eqa:apdx2:02} and performing several algebraic manipulations, we obtain the upper bound of queue backlogs as
\begin{align}
  & \epsilon T \sumk \bm 1^{\tt}\E_{\X}\cb{\q^{\A}\sq{k} + \q^{\U}\sq{k}} \nonumber\\
\le & \E_{\X}\cb{L\sq{0}} - \E_{\X}\cb{L\sq{K}} + TK\Psi + 2TKV\bar G \label{eqa:apdx2:03a}\\
\le & \E_{\X}\cb{L\sq{0}} + TK\Psi + 2TKV\bar G \label{eqa:apdx2:03b}
\end{align}
where \eqref{eqa:apdx2:03b} is due to the nonnegative $k$th frame Lyapunov function $\E_{\X}\cb{L\sq{K}}$.

Dividing both sides of \eqref{eqa:apdx2:03b} by $\epsilon TK$, we obtain
\begin{equation}\label{eqa:apdx2:04}
\begin{split}
 & \frac{1}{K}\sumk \bm 1^{\tt}\E_{\X}\cb{\q^{\A}\sq{k} + \q^{\U}\sq{k}} \\
\le&  \frac{\Psi + 2V\bar G }{\epsilon} + \frac{1}{\epsilon TK}\E_{\X}\cb{L\sq{0}}.
\end{split}
\end{equation}

Since the initial queue backlogs $\q^{\A}\sq{0}$ and $\q^{\U}\sq{0}$ are fixed, the initial Lyapunov function $\E_{\X}\cb{L\sq{0}}$ is bounded.
Letting $k \rightarrow \infty$, we obtain
\begin{equation}\label{eqa:apdx2:05}
\begin{split}
& \limsup\limits_{K\rightarrow \infty}\frac{1}{K}\sumk \bm 1^{\tt}\E_{\X}\cb{\q^{\A}\sq{k} + \q^{\U}\sq{k}} \\
\le & \frac{\Psi + 2V\bar G }{\epsilon} < \infty.
\end{split}
\end{equation}

The backlogs of access queues and processing queues are nonnegative due to the constraints in \eqref{eqa:12} and queue dynamic functions in \eqref{eqa:09} and \eqref{eqa:10}.
Based on the nonnegative queue backlogs and \eqref{eqa:apdx2:05}, we conclude that
\begin{equation}
\begin{split}
 &\limsup\limits_{K\rightarrow \infty}\frac{1}{K}\sumk \bm \E_{\X}\cb{q_{m,n}^{\A}\sq{k} + q_{m,n}^{\U}\sq{k}} \\
\le&  \frac{\Psi + 2V\bar G }{\epsilon} < \infty
\end{split}
\end{equation}
such that the constraints in \eqref{eqa:15} are satisfied.

Now, we prove the inequalities in \eqref{eqa:23}.
Based on \eqref{eqa:apdx2:05}, we obtain that the nonnegative queue backlogs of access queues and processing queues satisfy
\begin{equation}\label{eqa:apdx2:06}
\limsup\limits_{K\rightarrow \infty}\frac{1}{K}\sumk \bm 1^{\tt}\E_{\X}\cb{\q^{\A}\sq{k} } \le \frac{\Psi + 2V\bar G }{\epsilon} < \infty
\end{equation}
and
\begin{equation}\label{eqa:apdx2:07}
\limsup\limits_{K\rightarrow \infty}\frac{1}{K}\sumk \bm 1^{\tt}\E_{\X}\cb{\q^{\U}\sq{k}} \le \frac{\Psi + 2V\bar G}{\epsilon} < \infty.
\end{equation}

Based on \eqref{eqa:apdx2:06}, \eqref{eqa:apdx2:07}, and Theorem 2.8 in \cite{Neelybook}, we conclude that the access queues and processing queues are \mbox{mean-rate} stable.
Furthermore, the necessary conditions for mean-rate stable access queues and processing queues are obtained as \cite[Theorem 2.5]{Neelybook}
\begin{equation}\label{eqa:apdx2:08}
\bar\nu_{m,n} \le \limsup\limits_{K\rightarrow \infty} \frac{1}{KT} \sumk\sumtk \E_{\X}\cb{r_{m,n}\br{t_k}} \le \bar s_{m,n}.
\end{equation}

Hence, we obtain a relaxed LTGEE (R-LTGEE) minimization problem as
\begin{equation}\label{eqa:apdx2:09}
\begin{split}
G^* = \min\limits_{\Y} & \lim\limits_{K \rightarrow \infty} \frac{1}{KT}\sumk \sumtk \summ \E_{\X}\cb{ G_m^{\sg}\br{t_k} }\\
\st\; &  \eqref{eqa:06}, \eqref{eqa:12}-\eqref{eqa:14}   \mbox{ and }  \eqref{eqa:apdx2:08}
\end{split}
\end{equation}
where $G^*$ is the optimal value of the R-LTGEE minimization problem.

Since the constraints in \eqref{eqa:15} are a subset of the constraints in  \eqref{eqa:apdx2:08}, the optimal value of the LTGEE minimization problem \eqref{eqa:16} is lower-bounded by the optimal value of the R-LTGEE minimization problem \eqref{eqa:apdx2:09} as
\begin{equation}\label{eqa:apdx2:10}
G^* \le \lim\limits_{K \rightarrow \infty} \frac{1}{KT}\sumk \sumtk \summ \mathds{E}_{\X}\cb{ G_m^{\sg}\br{\Y^*\br{t_k}} }.
\end{equation}
Therefore, we establish the first inequality in \eqref{eqa:23}.

Based on the arguments in \cite{Neelybook}, when the random sources in $\X$ are stationary over different slots, there exists an optimal solution $\tilde\Y^* \triangleq \cb{\tilde\Y^*\br{t_k}}_{\forall t_k, k}$ to the R-LTGEE minimization problem \eqref{eqa:apdx2:09} that almost-surely satisfies: 1) $\tilde\Y^*\br{t_k}$ is a function of current random sources $\X\br{t_k}$; and 2) $\tilde\Y^*\br{t_k}$ guarantees that $\bar\nu_{m,n} \le \E_{\X}\cb{r_{m,n}\br{ \tilde\Y^*\br{t_k}  }} \le \bar s_{m,n}$ and $G^* = \sum\nolimits_{m=1}^M \E_{\X}\cb{ G_m^{\sg}\br{ \tilde\Y^*\br{t_k} } }$.

Since $\tilde\Y^* \triangleq \cb{\tilde\Y^*\br{t_k}}_{\forall k}$ is not a minimizer to the RHS of \eqref{eqa:apdx2:01b}, we obtain
\begin{align}
& \Delta_{\X}\sq{k} + V \sumtk\summ \mathds{E}_{\X}\cb{ G_m^{\sg}\br{\Y^*\br{t_k}} } \label{eqa:apdx2:12a}\\
\le & T\Psi + V \sumtk\summ \mathds{E}_{\X}\cb{ G_m^{\sg}\br{\tilde\Y^*\br{t_k}} } \nonumber\\
& + \sumtk \E^{\tt}_{\X}\cb{\v\br{t_k} - \r\br{\tilde\Y^*\br{t_k}}}\q^{\A}\sq{k} \nonumber\\
& + \sumtk \E^{\tt}_{\X}\cb{\r\br{\tilde\Y^*\br{t_k}} - \s\br{t_k}}\q^{\U}\sq{k} \label{eqa:apdx2:12b}\\
\le & T\Psi + TVG^* \label{eqa:apdx2:12c}
\end{align}
where the inequality \eqref{eqa:apdx2:12c} is based on the two facts: 1) $\tilde\Y^*\br{t_k}$ is a function of current random sources $\X\br{t_k}$; and 2) $\tilde\Y^*\br{t_k}$ guarantees that $\nu_{m,n} \le \E_{\X}\cb{r_{m,n}\br{ \tilde\Y^*\br{t_k}  }} \le s_{m,n}$ and $G^* = \sum\nolimits_{m=1}^M \E_{\X}\cb{ G_m^{\sg}\br{ \tilde\Y^*\br{t_k} } }$.

Taking telescoping summation over $k = 0, 1, \ldots, K-1$ over \eqref{eqa:apdx2:12c} and dividing both sides by $TKV$, we obtain
\begin{equation}\label{eqa:apdx:13}
\begin{split}
&\frac{1}{KT}\sumtk\summ \mathds{E}_{\X}\cb{ G_m^{\sg}\br{\Y^*\br{t_k}}} \\
& \le G^* + \frac{\Psi}{V} 
+ \frac{1}{2TKV}\norm{\q^{\A}\sq{0}}^2 
+ \frac{1}{2TKV}\norm{\q^{\U}\sq{0}}^2 .
\end{split}
\end{equation}

Letting $K \rightarrow \infty$, we obtain the second inequality of \eqref{eqa:23} due to the fixed backlogs $\q^{\A}\sq{0}$ and $\q^{\U}\sq{0}$.

\section{Proof of the Optimality of Scheduled UE Indicators in \eqref{eqa:27}}\label{apdx:03}
We prove the optimality of \eqref{eqa:27} via contradiction.

To minimize the RHS of \eqref{eqa:20} via allocating the scheduled UE indicator, we obtain the terms related to the scheduled UE indicators $a_{m,n}\sq{k}$ as
\begin{equation}\label{apdx3:01}
\begin{split}
&V \summ \E_{\chi}\cb{\sumtk G_m^{\sg}\br{t_k}} \\
&+ \summ\sumn\br{q_{m,n}^{\U}\sq{k} - q_{m,n}^{\A}\sq{k}}\E_{\chi}\cb{r_{m,n}\br{t_k}}
\end{split}
\end{equation}
where $G_m^{\sg}\br{t_k}$ and $r_{m,n}\br{t_k}$ are functions of $a_{m,n}\sq{k}$.

Based on the principle of opportunistically minimizing an expectation \cite{Neelybook}, the optimal scheduled UE indicators minimize
\begin{equation}\label{apdx3:02}
\begin{split}
& V\underbrace{\sumtk\summ  G_m^{\sg}\br{t_k}}_{\mbox{Grid-Energy Expenditure}} \\
& +  
\underbrace{\sumtk\summ\sumn\br{q_{m,n}^{\U}\sq{k} - q_{m,n}^{\A}\sq{k}}{r_{m,n}\br{t_k}}}_{\mbox{Data Rate of UEs}}.
\end{split}
\end{equation}
where $\frac{\partial G_m^{\sg}\br{t_k}}{\partial a_{m,n}\sq{k}} \ge 0$ is based on the definition of $G_m^{\sg}\br{t_k}$ in \eqref{eqa:08}.
Hence, the grid-energy expenditure monotonically increases with $a_{m,n}\sq{k}$.

Suppose that the $\br{m,n}$th UE is scheduled when $q_{m,n}^{\U}\sq{k} - q_{m,n}^{\A}\sq{k} \ge 0$ or $q_{m,n}^{\A}\sq{k} = 0$, namely $a_{m,n}\sq{k} = 1$ when $q_{m,n}^{\U}\sq{k} - q_{m,n}^{\A}\sq{k} \ge 0$ or $q_{m,n}^{\A}\sq{k} = 0$.
In the case with $q_{m,n}^{\U}\sq{k} - q_{m,n}^{\A}\sq{k} \ge 0$, we observe that data rate of the $\br{m,n}$th UE increases the value of \eqref{apdx3:02}.
To minimize \eqref{apdx3:02},  the data rate of the $\br{m,n}$th UE needs to be set to zero.
The data rate of the $\br{m,n}$th UE with $q_{m,n}^{\A}\sq{k} = 0$ needs to be set to zero following the similar arguments.
The scheduled UE indicator of the $\br{m,n}$th UE needs to be set as $a_{m,n} = 0$ which contradicts the assumption.
Therefore, we conclude that $a_{m,n}\sq{k} = 0$ when $q_{m,n}^{\U}\sq{k} - q_{m,n}^{\A}\sq{k} \ge 0$ or $q_{m,n}^{\A}\sq{k} = 0$.

Suppose that the $\br{m,n}$th UE is not scheduled when $q_{m,n}^{\U}\sq{k} - q_{m,n}^{\A}\sq{k} < 0$ and $q_{m,n}^{\A}\sq{k} \neq 0$, namely $a_{m,n}\sq{k} = 0$ when $q_{m,n}^{\U}\sq{k} - q_{m,n}^{\A}\sq{k} < 0$ and $q_{m,n}^{\A}\sq{k} \neq 0$.
Based on the previous reasoning, we obtain that the $\br{m,n}$th UE is not scheduled when $q_{m,n}^{\U}\sq{k} - q_{m,n}^{\A}\sq{k} \ge 0$ or $q_{m,n}^{\A}\sq{k} = 0$.
The $\br{m,n}$th UE will not be scheduled.
The backlog of access queue $q_{m,n}^{\A}\br{t_k}$ will become infinite which contradicts the $\br{m,n}$th queue-stable constraint in \eqref{eqa:15}.
Therefore, we conclude that $a_{m,n}\sq{k} = 1$ when $q_{m,n}^{\U}\sq{k} - q_{m,n}^{\A}\sq{k} < 0$ and $q_{m,n}^{\A}\sq{k} \neq 0$.

\section{Proof of the Activeness of Constraints in \eqref{eqa:33c}}\label{apdx:04}
Let $\tilde\Z\br{t_k} = \{\tilde\w_{m,n}\br{t_k}, \tilde\delta_{m}^{l}\br{t_k}\}$ denote the set of optimal beamforming vectors and exchanged NRE variables given $\phi\br{t_k}$.
Suppose that the $\br{m,n}$th constraint in \eqref{eqa:33c} is inactive, i.e.,
\begin{equation}\label{eqa:apdx3:01}
\frac{\h_{m,n}^{\H}\br{t_k}\tilde\w_{m,n}\br{t_k}}{f_{m,n}\br{\phi\br{t_k}}} > \sqrt{  I_{m,n}^{\intra}\br{t_k} +  I_{m,n}^{\inter}\br{t_k} + \sigma_{m,n}^2 }.
\end{equation}

Hence, we introduce an auxiliary variable $\gamma_{m,n}$ such that
\begin{equation}\label{eqa:apdx3:02}
\gamma_{m,n}\frac{\h_{m,n}^{\H}\br{t_k}\tilde\w_{m,n}\br{t_k}}{f_{m,n}\br{\phi\br{t_k}}} = \sqrt{  I_{m,n}^{\intra}\br{t_k} +  I_{m,n}^{\inter}\br{t_k} + \sigma_{m,n}^2 }.
\end{equation}

Based on \eqref{eqa:apdx3:01} and \eqref{eqa:apdx3:02}, we obtain $\gamma_{m,n} < 1$.
Hence, we can construct a new solution as $\bar\Z\br{t_k} = \{\bar\w_{m,n}\br{t_k}, \bar\delta_{m}^{l}\br{t_k}\}$ with $\bar\w_{m,n}\br{t_k} = \gamma_{m,n}\tilde\w_{m,n}\br{t_k}$ and $\bar\delta_{m}^{l}\br{t_k} = \tilde\delta_{m}^{l}\br{t_k}$.
We observe that the new solution $\bar\Z\br{t_k}$ satisfies all the constraints in \eqref{eqa:33b}--\eqref{eqa:33e}.
Moreover, the new solution $\bar\Z\br{t_k}$ obtains a smaller objective value than that of $\tilde\Z\br{t_k} = \{\tilde\w_{m,n}\br{t_k}, \tilde\delta_{m}^{l}\br{t_k}\}$ since $\gamma_{m,n}< 1$.
This observation contradicts with the assumption.
Hence, we conclude that the constraints in \eqref{eqa:33c} are active.

\bibliographystyle{IEEEtran}
\bibliography{dyj_bib}

\end{document}